\documentclass[journal]{IEEEtran}
\usepackage{cite}
\usepackage{amsmath,amssymb,amsfonts}
\usepackage{algorithm, algorithmic}
\usepackage{graphicx}
\usepackage{textcomp}
\usepackage{subcaption}
\usepackage{hyperref}
\usepackage{bm}

\newtheorem{theorem}{Theorem}
\newtheorem{lemma}{Lemma}
\newtheorem{proof}{proof}

\newtheorem{assumption}{Assumption}

\begin{document}
\title{Local Updates in Distributed Optimization: Provable Acceleration and Topology Effects}
\author{Zuang Wang and Yongqiang Wang, ~\IEEEmembership{Senior Member,~IEEE}
\thanks{Part of the work has been submitted to IEEE Conference on Decision and Control 2026  \cite{wang2026provableaccelerationdistributedoptimization}.}
\thanks{The work was supported in part by the National Science Foundation under Grants CCF-2106293, CCF-2215088, CNS-2219487, CCF-2334449, and CNS-2422312 (Corresponding author: Yongqiang Wang).} \thanks{The authors are with the Department of Electrical and Computer Engineering, Clemson University, Clemson, SC 29634, USA (e-mail: zuangw@clemson.edu; yongqiw@clemson.edu).} }

\maketitle

\begin{abstract}
Inspired by the success of performing multiple local optimization steps between communication rounds in federated learning, incorporating such local updates into distributed optimization has recently attracted growing interest.   However, unlike federated learning—where local updates can accelerate training by reducing gradient estimation error under mini-batch settings—it remains unclear whether similar benefits persist when exact gradients are available.  Moreover, existing theoretical results typically require reducing the step size when multiple local updates are employed, which can entirely offset any potential benefit of these additional local updates. In this paper, we focus on the classic DIGing algorithm and leverage the tight performance bounds provided by Performance Estimation Problems (PEP) to show that incorporating local updates can indeed accelerate distributed optimization. To the best of our knowledge, this is the first rigorous demonstration of such acceleration for a broad class of objective functions. Our analysis further reveals that, under an appropriate step size, performing only two local updates is sufficient to achieve the maximal possible improvement, and that additional local updates provide no further gains. Because more updates increase computational cost, these findings offer practical guidance for efficient implementation. We also show that these speed gains depend critically on the network structure, with sparser or less connected graphs—characterized by the spectral properties of the mixing matrix—yielding smaller improvements. Extensive experiments on both synthetic and real-world datasets corroborate the theoretical findings.
\end{abstract}


\section{Introduction}
In many systems, computational resources and data are inherently decentralized across multiple agents, such as autonomous robots, sensor networks, and modern machine learning architectures. This decentralization makes distributed optimization algorithms essential for coordinating computation and achieving global objectives \cite{ nedic2009distributed, predd2009collaborative, ren2008distributed}. We consider the following distributed optimization problem over an agent set $\mathcal{S}=\{1,2,\cdots, N\}$ as follows:
\begin{align}\label{problem}
\min_{x\in\mathbb{R}^d}f(x)=\frac{1}{N}\sum^{N}_{i=1}f_i(x),
\end{align}
where $f_i:\mathbb{R}^d\rightarrow\mathbb{R}$ denotes the local loss function, which is accessible only to agent $i\in\mathcal{S}$. In this setup, all agents collaboratively seek the minimizer $x^{\ast}$ to minimize the average of the agents’ objectives. Most existing distributed algorithms for solving this problem follow a “one-update, one-communication” pattern \cite{nedic2010constrained,  Shi2015EXTRA, jinminxu2015, sunsonata}, i.e., each local update is immediately followed by a communication step among agents. 

In recent years, motivated by the success of federated learning in using multiple local updates to accelerate convergence \cite{FedPAQ, li2020convergencefedavgnoniiddata}, several distributed optimization algorithms have been proposed \cite{Alghunaim2024Local, Ghiasvand2025Robust, Huang2025Distributed, Liu2025Guaranteeing} that allow a participating agent to perform multiple local updates before communicating with its neighbors. It is generally believed that such infrequent communication can reduce the total number of communication rounds required to achieve a given level of optimization accuracy.

However, without exception, all of these approaches require reducing the stepsize as the number of local updates increases, making it unclear whether local updates are truly beneficial. Federated learning benefits from multiple local updates because it assumes that gradients are computed using only mini-batches of samples \cite{woodworth2020local}; thus, increasing the number of local updates allows more data to be processed between communication rounds, thereby improving the quality of gradient estimation. In contrast, in the deterministic setting where exact gradients are available, this rationale no longer holds. Indeed, in the absence of gradient noise, simulation results in \cite{Alghunaim2024Local, performanceofGT, wu2025effectiveness} show that increasing the number of local updates yields almost no improvement when data heterogeneity is high or the communication graph is sparse. Moreover, simulation results in these papers suggest that increasing the number of local updates beyond certain thresholds provides only marginal benefits. Consequently, it remains unclear whether local updates are consistently beneficial, and if so, which factors govern their effectiveness.

In addition, most existing work on distributed optimization with multiple local updates suffers from the following two drawbacks:
\begin{enumerate}
    \item Existing works fail to provide rigorous {\it theoretical evidence} supporting the benefits of employing local updates. The analytical convergence rates reported in prior studies \cite{Alghunaim2024Local, Liu2025Guaranteeing, wu2025effectiveness, nguyen2023performance, proxskip} are derived from conservative analytical upper bounds rather than exact characterizations. More importantly, most of these bounds typically require the step size to decrease as the number of local updates $\tau$ increases. Consequently, although local updates reduce communication frequency, the enforced reduction in step size slows per-iteration progress, offsetting or even negating the potential gains from additional local updates. \cite{proxskip} shows the benefits of multiple local updates by comparing a multiple-local-update-algorithm ProxSkip with gradient descent using the same step size $\frac{1}{L}$ for both algorithms. However, this comparison may still be unfair, as the chosen step size can be suboptimal for either method and therefore may not reflect their respective best achievable performance. In summary, existing analyses are unable to explicitly demonstrate a theoretical advantage of employing local updates.
    \item Most of the existing works do not perform a controlled assessment of the effect of local updates in their {\it experimental comparisons}.  For example, \cite{kgt} compares results for different numbers of local updates using the same step size. Such a comparison may not accurately reflect the potential benefits of local updates, since algorithms with fewer local updates can typically adopt a larger step size. Consequently, fixing the step size places methods with fewer local updates at a disadvantage. A more balanced and informative comparison would instead use the optimal step size for each choice of the number of local updates, thereby evaluating their best achievable performance.
\end{enumerate}

In this work, leveraging the exact performance bounds provided by Performance Estimation Problems (PEP) \cite{pepsdp}, we systematically investigate whether local updates can indeed accelerate distributed optimization under exact gradients. Unlike traditional analytical approaches, which typically yield conservative asymptotic convergence bounds, PEP formulates the performance characterization problem as an optimization problem and provides \textbf{exact} performance bounds \cite{pepsdp}. By ``exact,'' we mean that the bound characterizes the true worst-case performance of the algorithm over the entire function class considered. In particular, one can explicitly construct a function within this class such that, when the algorithm is applied to it, the resulting optimization error attains the worst-case value predicted by the PEP and cannot be worse. This contrasts with analytical upper bounds, which may be conservative and potentially misleading \cite{pepsdp, meunier2025several}. To the best of our knowledge, this work is the first to employ the PEP framework to analyze the convergence speed of distributed optimization algorithms with local updates. Also, we emphasize that {\bf our PEP-based results are formally valid and mathematically rigorous}: they provide the exact worst-case performance of optimization algorithms over a {\bf prescribed function class} \cite{pepsdp}. This contrasts with conventional analyses, which typically rely on analytical upper bounds that may be loose, overly conservative, and potentially misleading \cite{meunier2025several}. It is also fundamentally different from numerical simulations conducted on specific function instances. {\bf Consequently, the conclusions derived from our PEP analysis are theoretically justified and rigorous}.

 We focus on the classical DIGing algorithm \cite{Nedic2017Geometric}, a representative gradient-tracking method that achieves exact convergence under a constant step size. Crucially, its convergence guarantees extend to time-varying mixing matrices, naturally enabling the incorporation of local updates (by setting the mixing matrix to the identity during local computation) without sacrificing convergence to the global optimum. Many existing distributed optimization methods (e.g., Aug-DGM \cite{jinminxu2015}, \cite{qu2017harnessing}, AsynDGM \cite{xu2017convergence}, AB \cite{xin2018linear}, Push-Pull \cite{Jinming_Xu1}, and NEXT \cite{di2016next}) can be viewed as variants of DIGing, making it a representative framework for studying the impact of local updates.
 
 The main contributions of this paper are summarized as follows \footnote{The first two contributions were discussed in the preliminary version \cite{wang2026provableaccelerationdistributedoptimization}, whereas the third contribution is new to this paper.}: 
\begin{enumerate}
    \item We rigorously prove that employing local updates can indeed accelerate the convergence of distributed optimization when exact gradients are available by using the \emph{exact} convergence bound provided by PEP. To the best of our knowledge, this theoretical result has not been reported previously. Furthermore, our analysis shows that performing just two local updates is sufficient to achieve the maximal improvement when an appropriate step size is chosen, and that additional local updates provide no further gains. Since extra local updates increase computational cost, our findings show that, with exact gradients, performing more than two local updates is unnecessary, providing a practical guideline for efficient distributed optimization.
    \item 
   We adapt the original PEP formulation for distributed algorithms proposed in \cite{pep2023tac} to our analysis setting. In particular, we extend it to incorporate boundedness constraints on the optimal solutions, which are common in practical distributed optimization problems. We also revise the formulation to reduce computational complexity, making it possible to analyze multiple local updates, which would otherwise be prohibitively expensive since existing PEP-based approaches are already costly for a single local update.

\item Leveraging PEP, we systematically study how the communication graph topology and the spectral properties of the mixing matrix influence the speed gains from local updates. Specifically, our results show that local updates generally yield smaller speed gains on sparser graphs, corroborating the numerical experiments in \cite{performanceofGT}. Moreover, our study shows that even with a fixed graph topology, reducing connectivity by adjusting the mixing matrix weights also diminishes the speed gains from multiple local updates.
\end{enumerate}

\section{A Motivating Example}  
In this section, we present an example demonstrating that the impact of the number of local updates on distributed optimization is far from intuitive under exact gradients, highlighting the need for careful theoretical investigation.

We consider distributed optimization with two agents, where the local objective functions are given by $f_1(x) = 0.5(x + 1)^2$ and $f_2(x) = 0.1(x - 1)^2$. One can verify that the corresponding smoothness constants are $L_1=1$ and $L_2=0.2$. Note that we intentionally choose $L_1$ and $L_2$
to be different so that the global optimal solution does not coincide with the average of the local optimal solutions. This avoids a trivial scenario in which each agent could simply perform local optimization independently and then average the final results \cite{kuwaranancharoen2018location, zamani2024set, kuwaranancharoen2024minimizer}. We initialize both agents at the average of the local optima, i.e., $x=\frac{x_1^\star+x_2^\star}{2}=0$, where $x_1^{\star}$ and $ x_2^{\star}$ denote the local optima of the two local objective functions, respectively. As shown in \cite{zamani2024set}, this initialization typically lies within the possible set of the global minimizers and is straightforward to tune in practice, since it only requires each agent to compute its own local optimum independently.
\begin{figure}[ht]
    \centering
    \includegraphics[width=0.8\columnwidth]{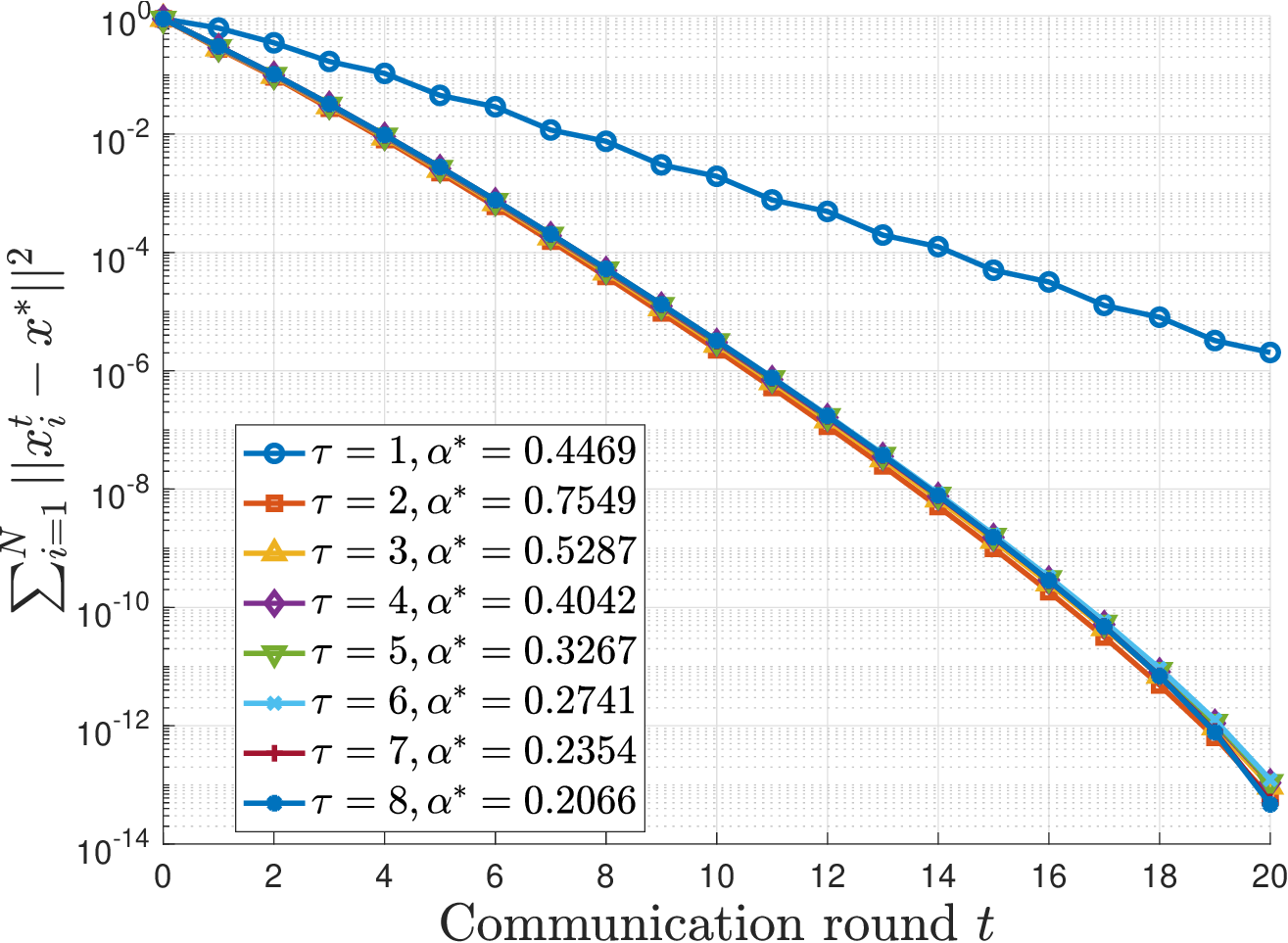}
    \caption{Evolution of the optimization error for different numbers of local updates ($\tau$) under their respective optimal step sizes ($\alpha^\ast$). The optimal step sizes were obtained via a grid search with a resolution of 0.0001. The two objective functions are $f_1(x) = 0.5(x + 1)^2$  and  $f_2(x) = 0.1(x - 1)^2$, with both agents initialized at zero. The mixing matrix is
$\begin{bmatrix} 0.5 & 0.5 \\ 0.5 & 0.5 \end{bmatrix}.$
    }
    \label{fig:motivation}
\end{figure}

The evolution of the optimization error for different numbers of local updates ($\tau$) under their respective optimal step sizes is illustrated in Figure~\ref{fig:motivation}. Interestingly, the results show that, with a properly tuned step size, incorporating local updates yields noticeable benefits over a single local update; however, $\tau = 2$ already achieves the maximal improvement. This raises a natural question: Does distributed optimization with multiple local updates indeed accelerate convergence, with the maximal improvement occurring at $\tau = 2$? And what factors potentially affect the speed gain? To answer this question, we employ exact worst-case performance bounds provided by PEP to conduct a rigorous and systematic comparison of DIGing under different numbers of local updates, thereby providing a theoretical answer.

\section{System Model and Algorithm Description}
\subsection{Notations}\label{notations}
We use $\mathbf{0}$ to denote matrices or vectors of appropriate dimensions with all elements equal to zero, and $\mathbf{1}$ to denote vectors of appropriate dimensions with all elements equal to one. We use $\otimes$ to denote the Kronecker product and $\mathcal{S}$ to denote the set $\{1, 2, \ldots, N\}$. For the convenience of presentation, we define the augmented state as
$
{\bm x}^k = [x^k_1,\; x^k_2,\; \dots,\; x^k_N] \in \mathbb{R}^{d \times N},
$
where the superscript $k$ belongs to the index set $I_K = \{0, \dots, K\}$.  We also define the augmented local optima and global optimum as
$
{\bm x}^\star = [x^\star_1,\; x^\star_2,\; \dots,\; x^\star_N]$ and $ {\bm x}^\ast = [x^\ast,\; x^\ast,\; \dots,\; x^\ast],
$
respectively.  
In order to streamline the notation, we augment the index set $I_K$ by including the two optimal states, resulting in the extended index sets $I_K^{\star} = \{0, \dots, K, \star\}$ and $I_K^{\star,\ast} = \{0, \dots, K, \star, \ast\}$. For example, $\{x^k\}_{k\in I^{\star,\ast}_K}$ denotes the set $\{x_0, ..., x^K, x^{\star}, x^{\ast}\}$.

Analogously, we define the augmented gradients and function values as
\begin{equation*}
\label{stacked_vector}
\begin{aligned}
&{\bm g}^k = [g^k_1,\; g^k_2,\; \dots,\; g^k_N] \in \mathbb{R}^{d \times N},\\ &{\bm f}^k = [f^k_1,\; f^k_2,\; \dots,\; f^k_N] \in \mathbb{R}^{N},\quad \text{for } k \in I_K^{\star,\ast}.
\end{aligned}
\end{equation*}




\subsection{Assumptions}

\begin{assumption}[Mixing Matrix]\label{ass:mixing_matrix}
Let \( G = (V, E) \) be an undirected and connected graph with \( |V| = N \) nodes. A matrix \( W \in \mathbb{R}^{N \times N} \) is said to be a mixing matrix associated with the graph \( G \) if it is symmetric (\( W = W^\top \)), doubly stochastic (\( W\mathbf{1} = \mathbf{1} \)), has entries satisfying \( 0 \leq W_{ij} \leq 1 \) for all \( i,j \in \mathcal{S} \), and respects the graph topology in the sense that \( W_{ij} > 0 \) whenever  \( (i,j) \in E \cup \{(i, i)\} \) and \( W_{ij} = 0 \) otherwise.
\end{assumption}

We assume that  every loss function $f_i(x)$  satisfies the following assumptions:
\begin{assumption}[Lipschitz Smoothness]\label{smooth_assumption}
Every $f_i(x)$ is $L$-smooth over $\mathbb{R}^d$, that is, there exists a constant $L>0$ such that
$$
\Vert \nabla f_i(x)-\nabla f_i(y)\Vert \leq L \Vert x-y\Vert
$$
holds for any $x,y\in\mathbb{R}^d$.
\end{assumption}

\begin{assumption}[Strong Convexity]\label{strong_convex_assumption}
Every $f_i(x)$   is $\mu$-strongly-convex over $\mathbb{R}^d$, that is, there exists a constant $\mu>0$ such that
$$
\langle\nabla f_i(x)-\nabla f_i(y),x-y\rangle \geq \mu\Vert x-y\Vert^2
$$
holds for any $x,y\in\mathbb{R}^d$.
\end{assumption}

Assumptions~\ref{smooth_assumption} and~\ref{strong_convex_assumption} ensure that the global loss function \(f(x)\) admits a unique minimizer \(x^\ast = \arg\min_{x \in \mathbb{R}^d} f(x)\). Moreover, for each $i$, the local function \(f_i\) admits a unique minimizer \(x_i^\star\). In other words, the global optimal solution $x^*$ satisfies 
\begin{equation}\label{global_optima}
    \frac{1}{N}\sum_{i=1}^N \nabla f_i(x^{\ast}) = \mathbf{0},
\end{equation}
and for each \(i\in\mathcal{S}\)
\begin{equation}\label{local_optima}
    \nabla f_i(x^{\star}_i) = \mathbf{0}.
\end{equation}
The clients have to share  information with each other to ensure that all clients can converge to the global optimal solution $x^*$.

\subsection{Incorporating Local Updates in DIGing}

In recent years, DIGing \cite{Nedic2017Geometric} and its many equivalents or variants of gradient-tracking algorithms (e.g., Aug-DGM \cite{jinminxu2015, qu2017harnessing}, AsynDGM \cite{xu2017convergence}, AB \cite{xin2018linear}, Push-Pull \cite{Jinming_Xu1}, and NEXT \cite{di2016next}) have garnered significant attention in distributed optimization because they can achieve exact convergence with a fixed step size. We focus on DIGing because it has guaranteed convergence under time-varying mixing matrices. This property naturally allows incorporating local updates (by setting the mixing matrix to the identity during local update steps) without sacrificing convergence to the exact global optimum. The DIGing algorithm with local updates is summarized in Algorithm~\ref{alg:diging_local}.

\begin{algorithm}[t]
\caption{DIGing with Local Updates}
\label{alg:diging_local}
\begin{algorithmic}[1]
\REQUIRE Initial point $x^0_i = \mathbf{0}\in \mathbb{R}^d$,
$y^0_i = g_i^0$,
step size $\alpha > 0$, the number of  local updates  $\tau \in \mathbb{Z}_{+}$, and the number of communication rounds  $T\in\mathbb{Z}_{+}$ (total number of iterations $K=T\times \tau$)
\FOR{$k = 0, 1, \ldots, K-1$}
    \IF{$k \in  \{\tau-1, 2\tau-1, 3\tau-1, \ldots\, K\tau-1\}$}
        \STATE \textbf{Communication round:}
        \[
        {\bm x}^{k+1} = W {\bm x}^k - \alpha {\bm y}^k 
        \]
        \[
        {\bm y}^{k+1} = W {\bm y}^k + {\bm g}^{k+1} - {\bm g}^k
        \]
    \ELSE
        \STATE \textbf{Local updates:}
        \[
        {\bm x}^{k+1} = {\bm x}^k - \alpha {\bm y}^k
        \]
        \[
        {\bm y}^{k+1} = {\bm y}^k + {\bm g}^{k+1} - {\bm g}^k
        \]
    \ENDIF
\ENDFOR
\end{algorithmic}
\end{algorithm}

\section{PEP-Based Characterization Showing That Local Updates Accelerate Distributed Optimization}
The pursuit of exact performance measures for optimization algorithms—rather than conservative asymptotic analyses—originated with \cite{nesterov2004applied} and was formalized by \cite{drori2014performance} via the Performance Estimation Problem (PEP). PEP formulates worst-case performance estimation as a convex semidefinite program (SDP) \cite{vandenberghe1996semidefinite}, which can be solved efficiently and exactly to yield  {\it exact} performance bounds over prescribed function classes, in contrast to the often loose guarantees provided by traditional asymptotic analyses \cite{pepsdp, meunier2025several}.

\subsection{Formulation}

We use $\mathcal{F}_{\mu, L}$ to denote the class of functions in which  each $f \in \mathcal{F}_{\mu, L}$ is $\mu$-strongly convex and $L$-smooth. 
Then we can formulate the PEP as follows:
\begin{align}
&\max_{{\substack{
    {\bm x}^0,\;{\bm g}^0,\;{\bm g}^1,\;\dots,\;{\bm g}^K,\;{\bm g}^{\star},\;{\bm g}^{\ast},\;{\bm x}^{\star},\;{\bm x}^{\ast}\\
    {\bm f}^0,\;{\bm f}^1,\;\dots,\;{\bm f}^K,\;{\bm f}^{\star},\;{\bm f}^{\ast}
}} 
} 
 \quad \frac{1}{N}\sum_{i=1}^N||x^K_i - x^*||^2\label{performance_measure}\\
\text{s.t.}\; & f_i\in\mathcal{F}_{\mu, L} \;\;\forall i \in \mathcal{S},\label{interpolation_constraints}\\
& \{x_{i}^k\},\;i\in \mathcal{S},\; k\in I_K\;\text{are generated by Algorithm}\;\ref{alg:diging_local},\label{algorithm_constraints}\\
& \textstyle\sum_{i=1}^N g_{i}^{\ast} = \textstyle\sum_{i=1}^N\nabla f_i(x^{\ast}) = 0,\;\text{corresponds to}\;\eqref{global_optima},\label{pep:global optimum}\\
& g^{\star}_{i} = \nabla f_i(x^{\star}_{i})=0,\;\forall i \in \mathcal{S},\;\text{corresponds to}\;\eqref{local_optima},\label{pep:local optimum}\\
& \|x_i^0 - x^{\ast}\|^2 \le R_0^2,\;\forall i\in \mathcal{S}, \|x_{i}^{\star} - x^{\ast}\|^2 \le R_{\ast}^2,\;\forall i \in \mathcal{S}.\label{last_two}
 \end{align}

Constraint \eqref{interpolation_constraints} guarantees that the function we considered in our PEP is in the corresponding functional class $\mathcal{F}_{\mu, L}$. Constraint (\ref{algorithm_constraints}) requires the local states to be compatible with Algorithm 1. Constraint  \eqref{pep:global optimum} ensures that $x^{\ast}$ is an optimal solution to problem \eqref{problem}, while \eqref{pep:local optimum} guarantees that each $x_i^{\star}$ is a local optimum of $f_i$. Conditions \eqref{last_two} encode the relationships between the initial states, the local minimizers, and the global minimizer, extending ideas introduced in \cite{pep2023tac}. 

It is worth noting that our PEP formulation differs from the existing PEP results for one-update-per-communication decentralized optimization in \cite{pep2023tac} in three key aspects. First, by periodically setting the mixing matrix to the identity matrix, our formulation enables the analysis of multiple local updates between communication rounds. In contrast, \cite{pep2023tac} is restricted to decentralized optimization with a single local update per communication under a time-invariant mixing matrix. Second, following recent works \cite{zamani2024set, kuwaranancharoen2024minimizer}, we impose more natural constraints \eqref{last_two} on the initial state, the local minimizers, and the global minimizer, whereas \cite{pep2023tac} achieved a similar effect by indirectly bounding the average error in the agents’ initial gradient estimates. Third, our formulation is designed to be as compact as possible, reducing the dimension of the resulting SDP variables by at least a factor of two compared with \cite{pep2023tac}. That is, we do not explicitly take the states $\{x_i^k\}$, for $i \in \mathcal{S}$ and $k = 1,\ldots, K$, generated by algorithm $\mathcal{M}$ from the initial states $\{x_i^0\}_{i \in \mathcal{S}}$, as decision variables in the PEP, nor do we introduce intermediate variables as in \cite{pep2023tac}. Instead, in \eqref{algorithm_constraints}, we express each $x_i^k$ as a linear combination of the initial states, which allows us to reformulate the PEP with fewer variables and constraints. This is crucial for our analysis, as solving PEPs is computationally expensive and becomes even more challenging when multiple local updates are incorporated.

The key challenge in making the aforementioned PEP a solvable SDP is that the constraints in \eqref{interpolation_constraints} are infinite-dimensional. Fortunately, this can be resolved with the following lemma:
\begin{lemma}(Interpolation for function class \(\mathcal{F}_{\mu, L}\)\cite{pepsdp})\label{lemma}
    Let \(I\) be any interpolation index set, and let \(\{(x^k, g^k, f^k)\}_{k\in I} \subset \mathbb{R}^d\times\mathbb{R}^d\times\mathbb{R}\). There exists \(f \in \mathcal{F}_{\mu, L}\) satisfying \(f(x^k) = f^k \) and \(g^k\in \partial f(x^k)\) for all \(k\in I\) if and only if the following inequality holds:
    \begin{equation}\label{interpolation_lemma}
    \begin{aligned}
     &  f_i(x_i) - f_j(x_j) - \langle g_j, x_i - x_j \rangle \geq \tfrac{1}{2}\!\left(1 - \tfrac{\mu}{L}\right)\times
         \\
      &   
        \left[ \tfrac{1}{L}\| g_i - g_j \|^2 + \mu \| x_i - x_j \|^2 
        - \tfrac{2\mu}{L} \langle g_j - g_i, x_j - x_i \rangle \right]
        \end{aligned}
    \end{equation}
\end{lemma}

 In this way, the constraints in (\ref{interpolation_constraints}) can be reformulated as equivalent, finite-dimensional constraints and thus as an equivalent SDP, making the problem solvable using standard SDP solvers. Thus, we have the following Theorem \ref{theorem_1}:

\begin{theorem}\label{theorem_1}
    Consider the class \(\mathcal{F}_{\mu, L}\) of \(L\)-smooth \(\mu\)-strongly convex functions, the performance criterion \( \frac{1}{N}\sum_{i=1}^N||x_i^K - x^{\ast}||^2\), and the fixed-step-size DIGing algorithm that computes \(K\) iterates according to mixing matrix \(W \in \mathbb{R}^{N \times N}\). The \textbf{exact} worst-case optimization error over the function class \(\mathcal{F}_{\mu, L}\) after \(K\) iterations is equal to the optimal value of the PEP problem characterized by (\ref{performance_measure})-(\ref{last_two}).
\end{theorem}
\begin{proof}
    See  Appendix \ref{proof of theorem 1}.
\end{proof}

\subsection{Results}
In this section, we present the results obtained by solving the PEP described above. We first specify the parameter settings used in the PEP. In particular, we set $\mu = 0.1$ and $L = 1$ for all local objective functions, and choose $R_0 = R_{\ast}$ following \cite{zamani2024set}.  Setting $R_0 = R_{\ast}$ is reasonable because local agents can effectively use local updates to bring their states close to their local optima. Consequently, local updates steer each state mainly toward its local optimum without significantly aiding the convergence of the average state to the global optimum, which complicates the identification of their impact on local updates.

To enable a fair comparison of the DIGing algorithm across different numbers of local updates, we perform a grid search over the step size for each setting to minimize the exact worst-case optimization error. Determining the optimal step size is a long-standing problem in optimization. Recently, the PEP methodology has been used to reformulate the problem of finding time-varying optimal step sizes for centralized gradient descent as a nonconvex quadratically constrained quadratic program (QCQP) \cite{2024bnbqcqp, kamri2025numericaldesignoptimizedfirstorder}. In \cite{2024bnbqcqp}, the author solves the nonconvex QCQP using a Branch-and-Bound method, which can be computationally intensive—especially in our setting with multiple local updates, where the number of iterations can be large. In \cite{kamri2025numericaldesignoptimizedfirstorder}, the authors propose several iterative methods for solving the nonconvex QCQP by exploiting its structural properties. However, due to the inherent nonconvexity of the problem, convergence to a global optimum---i.e., the optimal step size---is not guaranteed. 

\begin{figure}[ht]
\centerline{\includegraphics[width=0.8\columnwidth]{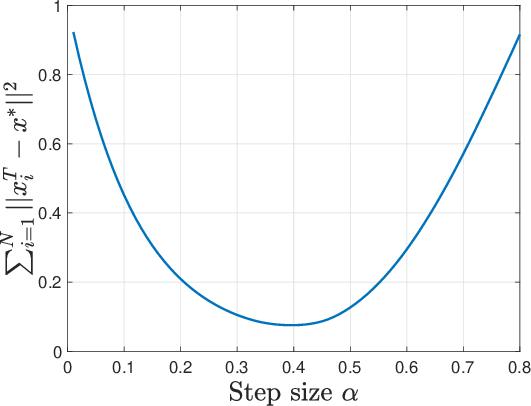}}
\caption{Algorithm \ref{alg:diging_local}’s exact worst-case convergence error at the final communication round $T$ (with $K = T\tau$ iterations), shown as a function of the step size for $\tau = 4$ local updates on an all-to-all graph. The step size is selected via a grid search over $\alpha \in [0.01, 0.8]$ with resolution $0.01$.}
\label{grid_search_demonstration}
\end{figure}

Fortunately, since Algorithm \ref{alg:diging_local} involves only a single fixed step size, it can be determined via a simple grid search, similar in spirit to the black-box optimization approach used in \cite{meunier2025several}. In our setting, however, performing a fine-grained grid search is more suitable, as accurately estimating the optimal step size is essential both for characterizing the best achievable performance of Algorithm~\ref{alg:diging_local} and for identifying how the optimal step size varies with the number of local updates. Figure~\ref{grid_search_demonstration} shows an example of our grid search result. It can be seen that for each number of local updates $\tau$, the optimal step size for Algorithm~\ref{alg:diging_local} is unique. The same observation holds as the number of local updates or the mixing matrix varies.

Based on this interpretation, for each number of local updates, we perform a grid search to identify the unique optimal step size for Algorithm~\ref{alg:diging_local}, enabling a fair comparison of the effect of the number of local updates by comparing their best achievable \emph{exact} convergence performance. Since solving the PEP is computationally intensive, we first restrict our experiments to a $4$-agent setting. We will show in the next section that similar results can be obtained for more agent scenarios. Despite this limitation, the considered setup still covers most common graph topologies in distributed optimization, including the all-to-all graph, the ring graph, and random graphs generated according to the Erdős--Rényi model $G(N, p)$  with edge probability \(p=0.6\) and weights following the Metropolis-Hastings rule \cite{pep2023tac}.  
\begin{figure*}[t]
\centering

\begin{subfigure}{\columnwidth}
    \centering
    \captionsetup{font=small}
    \includegraphics[width=0.8\linewidth]{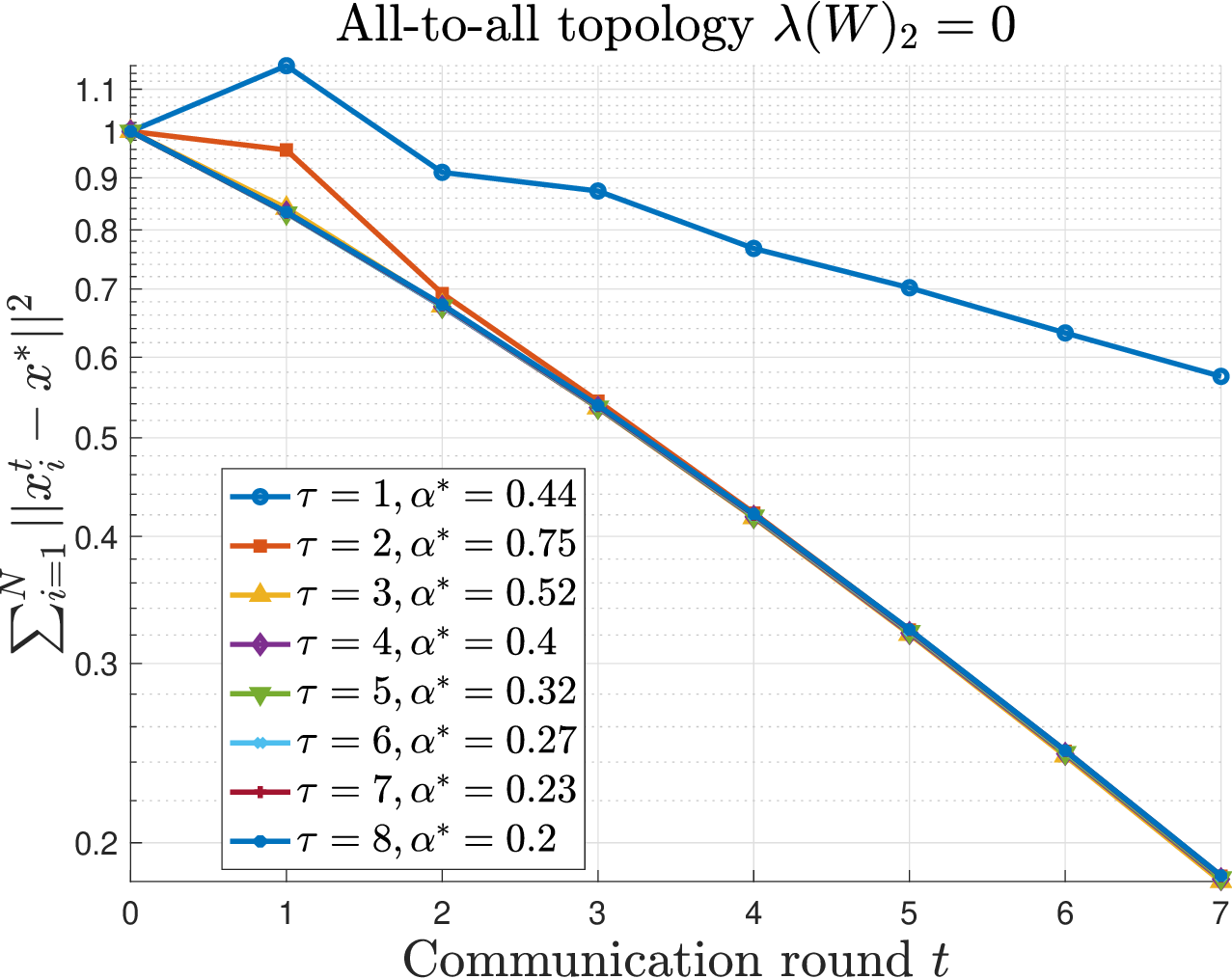}
    \caption{}
    \label{fig:alltoall_pep}
\end{subfigure}
\hfill
\begin{subfigure}{\columnwidth}
    \centering
    \captionsetup{font=small}
    \includegraphics[width=0.8\linewidth]{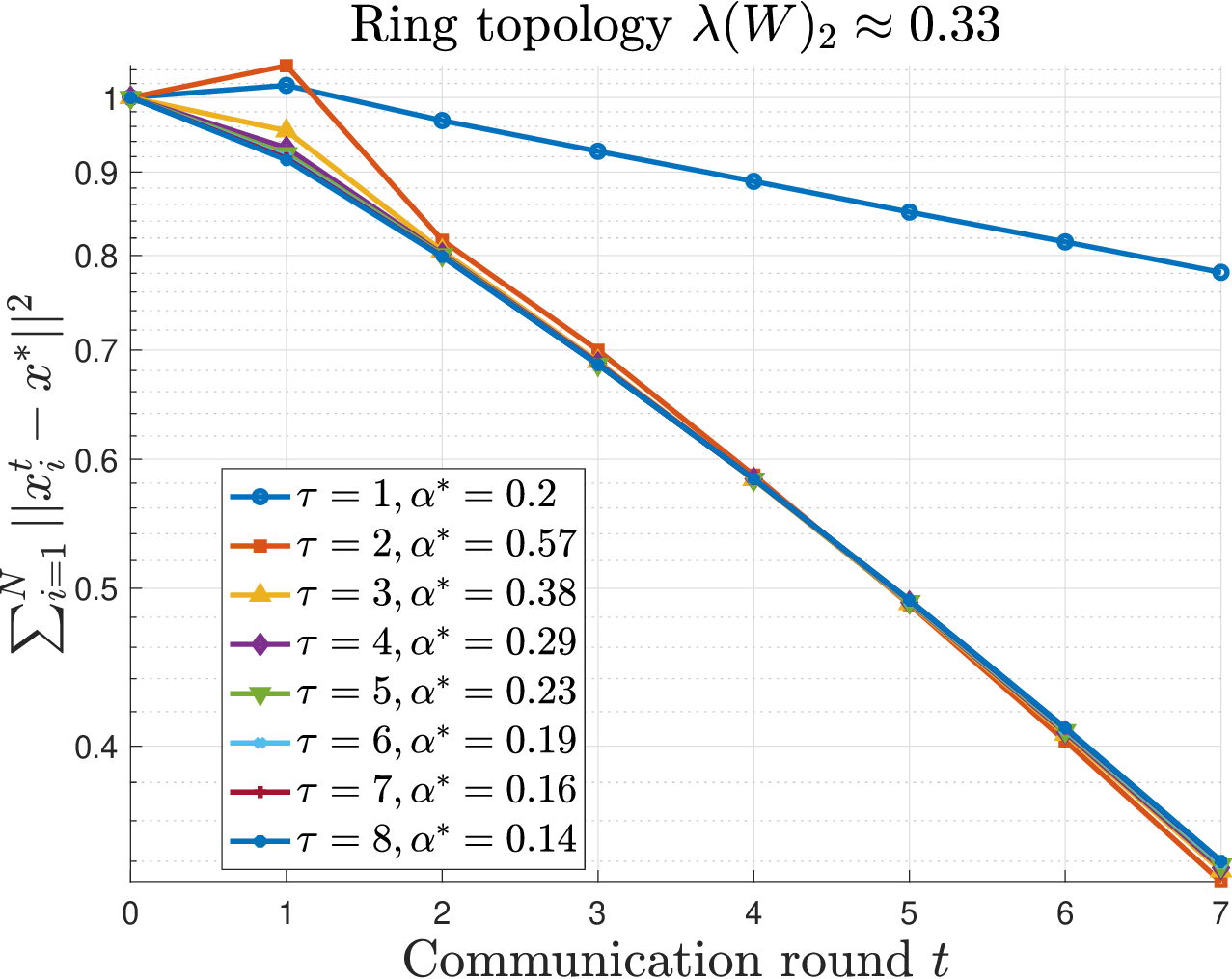}
    \caption{}
    \label{fig:ring_pep}
\end{subfigure}
\hfill
\begin{subfigure}{\columnwidth}
    \centering\
    \captionsetup{font=small}
    \includegraphics[width=0.8\linewidth]{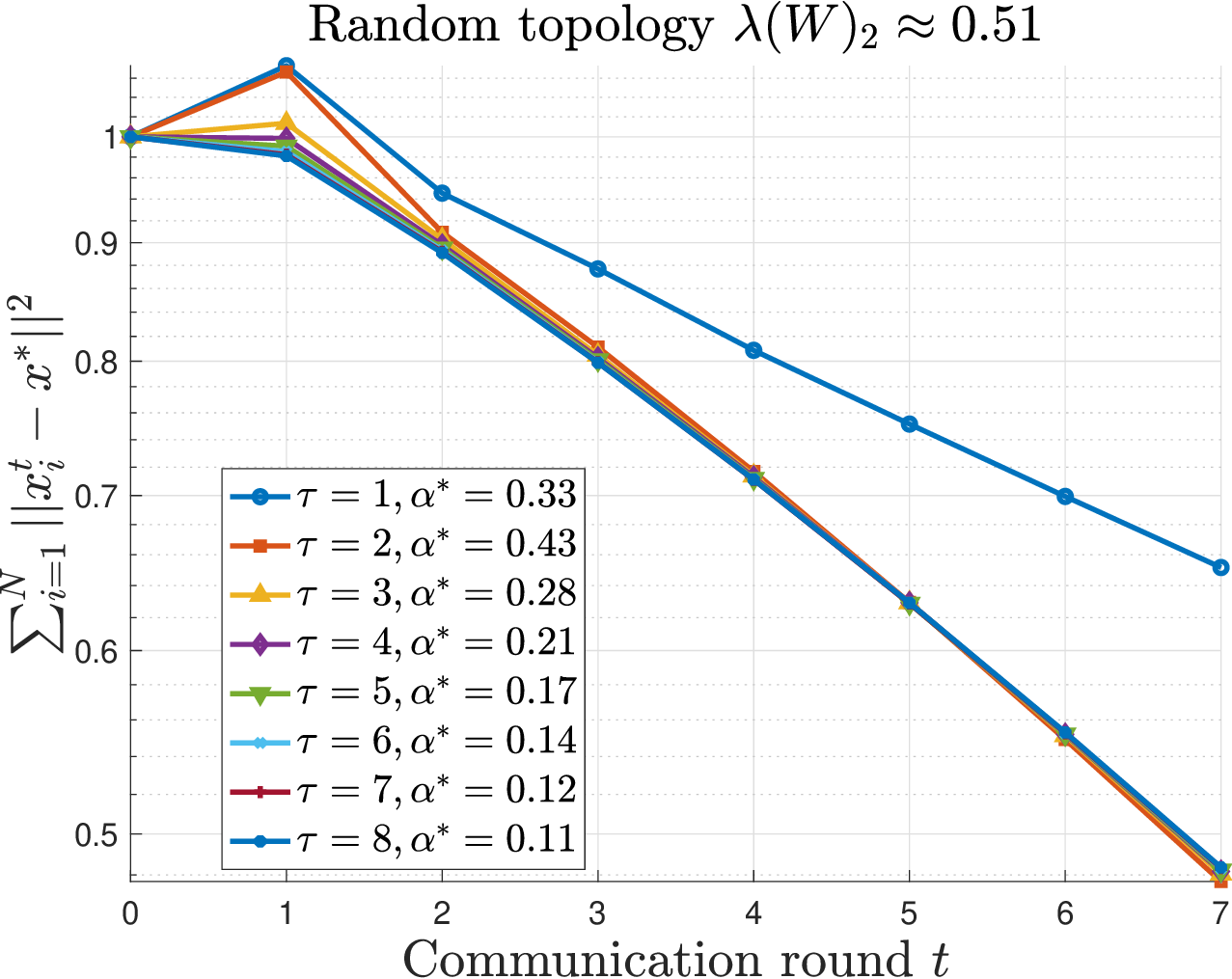}
    \caption{}
    \label{fig:random_pep}
\end{subfigure}
\begin{subfigure}{\columnwidth}
    \centering
    \captionsetup{font=small}
    \includegraphics[width=0.8\linewidth]{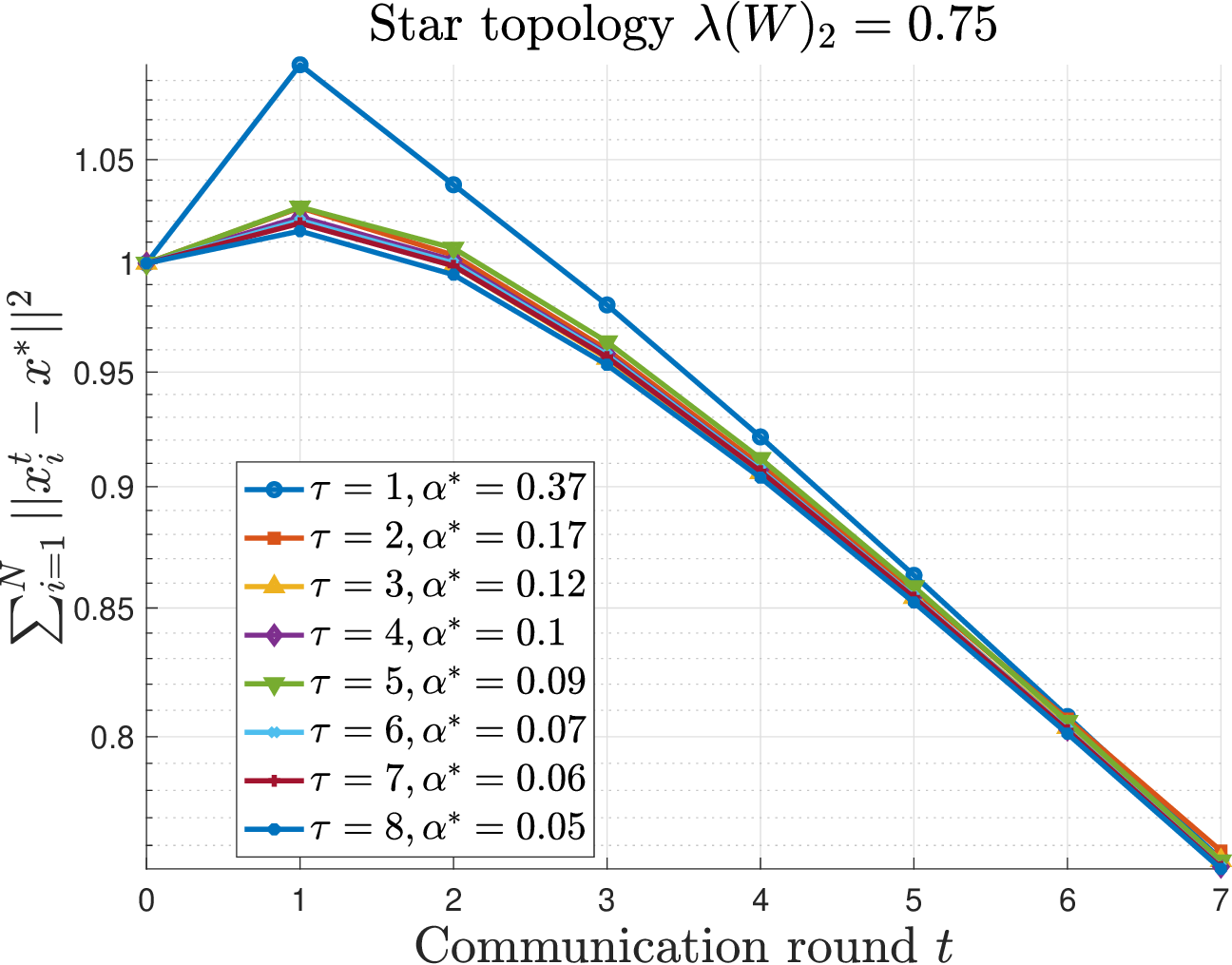}
    \caption{}
    \label{fig:star_pep}
\end{subfigure}
\caption{PEP-based exact worst-case optimization error of Algorithm~\ref{alg:diging_local} for different numbers of local updates, each using its corresponding optimal step size $\alpha^\ast$, across various graph topologies. For each $\tau$, $\alpha^\ast$ was determined via a grid search over $\alpha \in [0.01, 0.8]$ with 0.01 resolution for the function class $\mathcal{F}_{\mu, L}$ with $\mu = 0.1$ and $L = 1$.}
\label{fig:pep}
\end{figure*}
The main results are shown in Figure~\ref{fig:pep}, from which two key observations can be made:
\subsubsection{Local Updates Accelerate DIGing}Local updates consistently accelerate convergence in distributed optimization with exact gradients across different communication graphs, although the magnitude of the speedup varies with the graph topology. To the best of our knowledge, this is the first work to rigorously establish such results for a general class of functions. Moreover, our results indicate that the maximum speedup occurs at $\tau = 2$, and further increasing the number of local updates provides no additional improvement. This observation is particularly relevant in practice, as performing more local updates increases computational complexity. To the best of our knowledge, this is the first theoretical result establishing a saturation effect in the acceleration provided by local updates for distributed optimization. 
In the next section, we systematically investigate—using PEP results—how the effectiveness of local updates is influenced by the communication graph and mixing matrix.
\subsubsection{Optimal Step Size \texorpdfstring{$\alpha^{\ast} \propto 1/\tau$}{alpha* proportional to 1/tau}} When $\tau\geq2$, the optimal step size decreases as $\tau$ increases. This behavior is partially consistent with existing theoretical analyses, which typically require the step size to scale inversely with the number of local updates (see, e.g., $\alpha \le \mathcal{O}(1/\tau)$ as $\tau$ grows in \cite{Nedic2017Geometric, Alghunaim2024Local}). However, existing analyses only provide upper bounds on the admissible step size, whereas our results characterize the optimal step size that minimizes the exact worst-case performance. Interestingly, the optimal step size for $\tau = 2$ exceeds that for $\tau = 1$, a behavior not explained by existing theory but consistent with the motivating example in Figure~\ref{fig:motivation}. Moreover, when $\tau$ is sufficiently large, the optimal step size approximately follows $\alpha^{\ast} \propto 1/\tau$. This scaling behavior is consistent with the motivating example in Figure~\ref{fig:motivation}, and, as we will show in Sec. VI, it also persists in more complex regression problems. These results provide valuable guidance for choosing an effective step size in distributed optimization and learning.


\section{Network-Independent PEP Formulation and Mixing Matrix Effects on Local Update Speed Gain}
\subsection{Network-Independent PEP Formulation} 
The information encoded by the mixing matrix $W$ is captured through the constraints in \eqref{algorithm_constraints}, particularly via the iteration dynamics of Algorithm \ref{alg:diging_local}. Rather than fixing a specific $W$, we can consider a broader class of mixing matrices within the PEP framework and characterize the worst-case optimization error over this class. Specifically, the consensus steps can be reformulated as constraints in which the mixing matrix is treated as a decision variable subject to spectral constraints, rather than as an explicit $N \times N$ matrix.

\subsubsection{Reformulate the consensus steps in Algorithm~\ref{alg:diging_local}} The consensus step plays a central role in the PEP formulation for distributed algorithms. In the DIGing algorithm, each communication round consists of two consensus steps: one applied to the state variables $\{x_i^k\}$ and the other to the gradient tracking variables $\{y_i^k\}$. A key challenge in constructing a network-independent PEP is to accurately represent these consensus operations within the PEP constraints. Following \cite{pep2023tac, colla2025exploitingagentsymmetriesperformance}, we address this by treating the mixing matrix $W$ as a decision variable rather than a fixed matrix. This approach allows us to derive worst-case performance bounds over a broad class of mixing matrices, instead of restricting attention to a single fixed matrix. By encoding the consensus dynamics through spectral constraints on $W$, the resulting PEP formulation yields performance guarantees that hold uniformly over all communication graphs satisfying the prescribed spectral conditions. We can rewrite the consensus step in DIGing associated with $\{x_i^k\}$ and $\{y_i^k\}$ as follows, respectively for each $i\in \mathcal{S}$ and $k\in \{\tau-1, 2\tau-1, 3\tau-1, \ldots\, T\tau-1\}$ 
\begin{equation}\notag
    \tilde{x}_i^{k} = x_i^{k+1} + \alpha y_i^k = \sum_{j=1}^{N}w_{ij}x_j^k,
\end{equation}
\begin{equation}\notag
    \tilde{y}_i^{k} = y_i^{k+1} - g_i^{k+1} + g_i^{k} = \sum_{j=1}^{N}w_{ij}y_j^k.
\end{equation}
Accordingly, the consensus steps can be reformulated as constraints involving the aggregate variables $x_i^{k+1} + \alpha y_i^k$ and $y_i^{k+1} - g_i^{k+1} + g_i^{k}$ in relation to the primal and tracking states $x_j^k$ and $y_j^k$, respectively. Notably, because $\tilde{x}_i^{k}$ and $\tilde{y}_i^{k}$ represent linear combinations of existing variables, this approach obviates the need for auxiliary intermediate variables. Consequently, the resulting semidefinite program (SDP) remains low-dimensional, enabling an efficient PEP formulation.

We define $W(\lambda^{-}, \lambda^{+})$ as the set of mixing matrices with given eigenvalue bounds $\lambda^{-}, \lambda^{+} \in (-1, 1)$ (excluding $\lambda_1 = 1$):
$$\left\{ 
W \in \mathbb{R}^{N \times N} : 
\begin{aligned}
    & W \;\text{is defined in Assumption \ref{ass:mixing_matrix}}\\
    & \lambda^{-} \leq \lambda_N(W) \leq \dots \leq \lambda_2(W) \leq \lambda^{+}
\end{aligned}
\right\}$$
where $\lambda_n$ denotes the $n$-th largest eigenvalue. For any $W$ satisfying Assumption \ref{ass:mixing_matrix}, we have $\lambda_1(W) = 1$, with the corresponding eigenvector $\mathbf{1}_N$. We define the following stacked version of the state variables and gradient tracking variables for the communication-only steps as follows:
\begin{equation}\notag
    \begin{aligned}
        &\tilde{X}_i = [\tilde{x}_i^{\tau-1}\;\tilde{x}_i^{2\tau-1}\;... \;\tilde{x}_i^{T\tau-1} ],
        \; X_i = [x_i^{\tau-1}\;x_i^{2\tau-1}\;... \;x_i^{T\tau-1} ],\\
        &\tilde{Y}_i = [\tilde{y}_i^{\tau-1}\;\tilde{y}_i^{2\tau-1}\;... \;\tilde{y}_i^{T\tau-1} ],
        \; Y_i = [y_i^{\tau-1}\;y_i^{2\tau-1}\; ... \;y_i^{T\tau-1} ].
    \end{aligned}
\end{equation}
Then, we can further define $$
X = \begin{bmatrix}
X_1 \\
X_2 \\
\vdots \\
X_n
\end{bmatrix}
\in \mathbb{R}^{nd \times T}.
$$ The matrices $\tilde{X}$, $Y$, and $\tilde{Y}$ can be defined analogously. Additionally, we define 
\begin{equation}\notag
    \begin{aligned}
        &X_{avg} = [x_{avg}^{\tau-1}\;x_{avg}^{2\tau-1}\;... \;x_{avg}^{T\tau-1} ],\\
        &\tilde{X}_{avg} = [\tilde{x}_{avg}^{\tau-1}\;\tilde{x}_{avg}^{2\tau-1}\;... \;\tilde{x}_{avg}^{T\tau-1} ],\\
        &Y_{avg} = [y_{avg}^{\tau-1}\;y_{avg}^{2\tau-1}\; ... \;y_{avg}^{T\tau-1} ],\\
        &\tilde{Y}_{avg} = [\tilde{y}_{avg}^{\tau-1}\;\tilde{y}_{avg}^{2\tau-1}\;... \;\tilde{y}_{avg}^{T\tau-1} ],
    \end{aligned}
\end{equation}
where the average is taken over $i$; for example $\tilde{x}^{\tau}_{avg} = \frac{1}{N}\sum_{i=1}^N\tilde{x}^{\tau}_{i}$. The reformulation of the consensus steps is based on
Theorem 4 in \cite{colla2025exploitingagentsymmetriesperformance}. For
completeness, we restate it here using our notation.
\begin{lemma}(Theorem 4 in \cite{colla2025exploitingagentsymmetriesperformance})\label{lemma 2}
Let $I$ be an index set and let $\lambda^{-}, \lambda^{+} \in (-1,1)$ with $\lambda^{-} \le \lambda^{+}$.
There exists $W \in \mathcal{W}(\lambda^{-}, \lambda^{+})$ such that \(
\tilde{X} = (W \otimes I_d)X
\) if and only if the following inequalities hold:

\begin{equation}\notag
X_{//} = \tilde{X}_{//},
\end{equation}

\begin{equation}\notag
(\tilde{X}_{\perp} - \lambda^{-} X_{\perp})^{T}(\tilde{X}_{\perp} - \lambda^{+} X_{\perp})
\preceq 0,
\end{equation}

\begin{equation}\notag
X_{\perp}^{T} \tilde{X}_{\perp} = \tilde{X}_{\perp}^{T} X_{\perp},
\end{equation}
where $X_{//} = X_{avg} \otimes \mathbf{1}_n$, $X_{\perp} = X - X_{//}$, $\tilde{X}_{//} = \tilde{X}_{avg} \otimes \mathbf{1}_n$ and $\tilde{X}_{\perp} = \tilde{X} - \tilde{X}_{//}$.
\end{lemma}
By using this lemma, we can then reformulate the consensus step with respect to the state variables as follows:
\begin{equation}\label{state variables}
    \begin{aligned}
        & \left(\frac{1}{N} \sum_{i=1}^{N} (X_i - \tilde{X}_i)\right)^T \left( \frac{1}{N} \sum_{j=1}^{N} (X_j - \tilde{X}_j) \right) = 0, \\[10pt]
        & \frac{1}{N} \sum_{i=1}^{N} \Big[ (\tilde{X}_i - \tilde{X}_{avg}) - \lambda^{-} (X_i - X_{avg}) \Big]^T \dots \\
        & \quad \dots \Big[ (\tilde{X}_i - \tilde{X}_{avg}) - \lambda^{+} (X_i - X_{avg}) \Big] \preceq 0, \\[10pt]
        & \frac{1}{N} \sum_{i=1}^{N} (X_i - X_{avg})^T (\tilde{X}_i - \tilde{X}_{avg}) \\
        & \quad - \frac{1}{N} \sum_{i=1}^{N} (\tilde{X}_i - \tilde{X}_{avg})^T (X_i - X_{avg}) = 0.
    \end{aligned}
\end{equation}
The formulation of the communication steps for the gradient tracking variables follows an analogous derivation; we omit it here due to page constraints.

Since all the mixing matrices $W$ used in the consensus steps of 
Algorithm \ref{alg:diging_local} are reformulated through constraints 
involving $\lambda^{-}$ and $\lambda^{+}$, which together characterize the 
spectral properties of $W$, the original PEP formulation described by 
\eqref{performance_measure}--\eqref{last_two} can be reformulated as a network-independent PEP that returns the worst-case optimization error over a general class of mixing matrices $\mathcal{W}(\lambda^{-}, \lambda^{+})$ rather than a fixed $W$.
\subsubsection{Agent-Number Independence of the Network-Independent PEP} Note that after the consensus steps are reformulated, the resulting constraints exhibit permutation symmetry across all agents. In other words, this observation means that all the agents can be permuted in a PEP solution without impacting its worst-case value, and thus no agent plays a distinct role in the algorithm or its performance evaluation. We can therefore reformulate our network-independent PEP into an agent-independent PEP by using a similar technique as provided in \cite{colla2025exploitingagentsymmetriesperformance}.  Therefore, in practice, it suffices to use the results obtained from the two-agent network-independent PEP. This statement is formalized in the following theorem.
\begin{theorem}\label{2=n}
    Consider a set of local functions $\{f_i\}_{i\in\mathcal{S}}$ that all belong to the function class $\mathcal{F}_{\mu, L}$. Suppose Algorithm \ref{alg:diging_local} is applied and the mixing matrix is encoded in the PEP constraints as in Lemma \ref{lemma 2}. Then, the resulting PEP outputs are independent of the number of agents $N$. That is, for any $n\ge 2$, we have 
    \begin{equation}\notag
        w_2(\tau, \alpha, T) = w_n(\tau, \alpha, T),\; \forall n\ge 2,
    \end{equation}
    where $w_N(\tau, \alpha, T)$ denotes the exact worst-case optimization error for an $N$-agent distributed optimization problem after $T$ communication rounds, given $\tau$ local updates and step size $\alpha$.
\end{theorem}

\begin{proof} After applying Lemma \ref{lemma 2} to reformulate the communication round state variables $\{x_i^k\}$ for each $i \in \mathcal{S}$ and $k \in \{\tau-1, 2\tau-1, 3\tau-1, \ldots, T\tau-1\}$ into \eqref{state variables}, the resulting constraints \eqref{state variables} become independent of the specific agent index. The same property holds for the gradient tracking variables $\{y_i^k\}$. Consequently, all agents are equivalent, and any permutation of the agent indices leaves the PEP unchanged. Therefore, the solution of our PEP falls within the class of fully symmetric solutions as defined in Corollary 6.1 of \cite{colla2025exploitingagentsymmetriesperformance}. By Theorem 7 in \cite{colla2025exploitingagentsymmetriesperformance}, the resulting SDP of such a PEP can be reformulated as an equivalent agent-number-independent SDP. Then, it follows that our PEP results do not depend on the number of agents, and hence $w_2(\tau, \alpha, T) = w_n(\tau, \alpha, T)$ for all $n \ge 2$. 
\end{proof}

\subsection{Effects of Mixing Matrix on the Speed Gain from Local Updates}
In this section, we investigate the effect of the mixing matrix on the speed gain obtained from local updates. First, we define the speed gain as the ratio between the best achievable (under the optimal step size) worst-case optimization error when no local updates are performed ($\tau = 1$) and the best achievable worst-case optimization error when multiple local updates are performed ($\tau \ge 2$). Formally, the speed gain is defined as

\begin{equation}\label{eq:speed_gain}
    \frac{w_N(1, \alpha^*(\tau), T)}{w_N(\tau, \alpha^*(1), T)}, \quad \tau \ge 2.
\end{equation}

To the best of our knowledge, the influence of network topology on the convergence speed gains obtained by employing multiple local updates were only studied empirically in \cite{performanceofGT}. 
In contrast, we use PEP to provide a rigorous theoretical analysis over a broad class of functions.
\begin{figure}[ht]
    \centering
    \includegraphics[width=0.9\columnwidth]{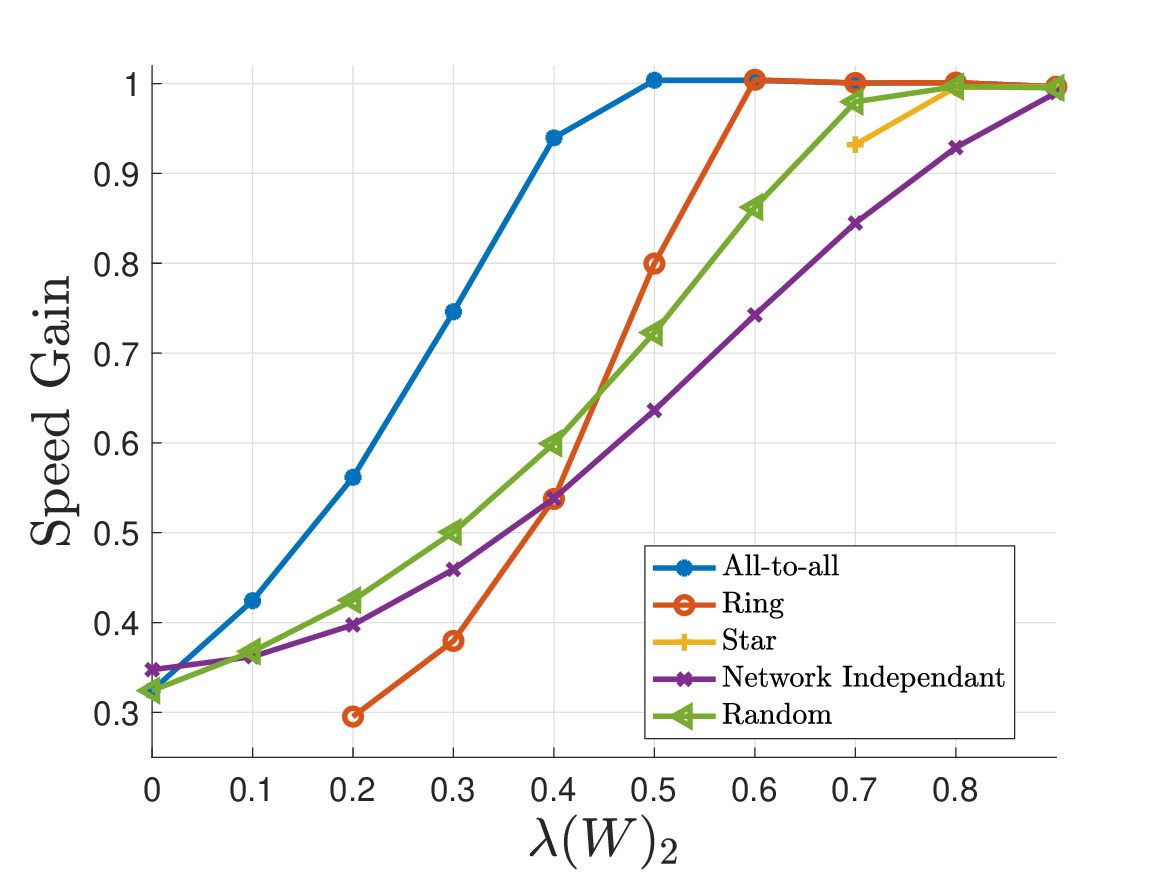}
    \caption{Speed gain of DIGing from multiple local updates as a function of the second-largest eigenvalue $\lambda_2(W)$ of the mixing matrix. The speed gain is defined as the ratio of the worst-case optimization error at communication round $T$ achieved with $\tau = 2$ and its corresponding optimal step size to that achieved with $\tau = 1$ under its optimal step size. To examine the influence of the network, we vary the weights of the 
    mixing matrix under different 4-agent fixed communication graphs, which leads 
    to different mixing matrices and corresponding values of $\lambda_2(W)$. We also directly vary $\lambda_2(W)$ in the network-independent PEP formulation.
    }
    \label{fig:exact_mixing_matrix_different_rho}
\end{figure}

We study the effects of the mixing matrix in the following two ways. First, we fix the communication graph and adjust the weights of the mixing matrix.
For example, when the communication graph is fixed as a 4-agent ring network, the mixing matrix has the following form:
\[
W =
\begin{bmatrix}
1-2a & a & 0 & a \\
a & 1-2a & a & 0 \\
0 & a & 1-2a & a \\
a & 0 & a & 1-2a
\end{bmatrix}
\]
Thus, we have the eigenvalues of $W$ as follows:
\[
\{\lambda_1,\lambda_2,\lambda_3,\lambda_4\}
=
\{1,\;1-2a,\;1-4a,\;1-2a\}.
\]
So by adjusting the value of $a$, we can get different $W$ with different $\lambda_2(W)$. Similar techniques apply to star graphs, ring graphs, and random graphs. Second, we change the parameters $\lambda^{-}, \lambda^{+}$ in the network-independent PEP formulation, where the mixing matrix is reformulated as constraints with respect to $\lambda^{-}$ and $ \lambda^{+}$. The results are shown in the Figure \ref{fig:exact_mixing_matrix_different_rho}. Across all considered scenarios, the achievable speed gain decreases as $\lambda_2(W)$ increases. This indicates that the convergence speed gain obtained by employing multiple local updates diminishes as the connectivity of the mixing matrix, quantified by $\lambda_2(W)$, worsens.

\section{Numerical Experiments}
In this section, we apply DIGing to practical learning tasks on both synthetic data and real-world data to evaluate the theoretical results obtained using PEP.
\begin{figure*}[t]
\centering
\begin{subfigure}{\columnwidth}
    \centering
    \captionsetup{font=small}
    \includegraphics[width=0.8\linewidth]{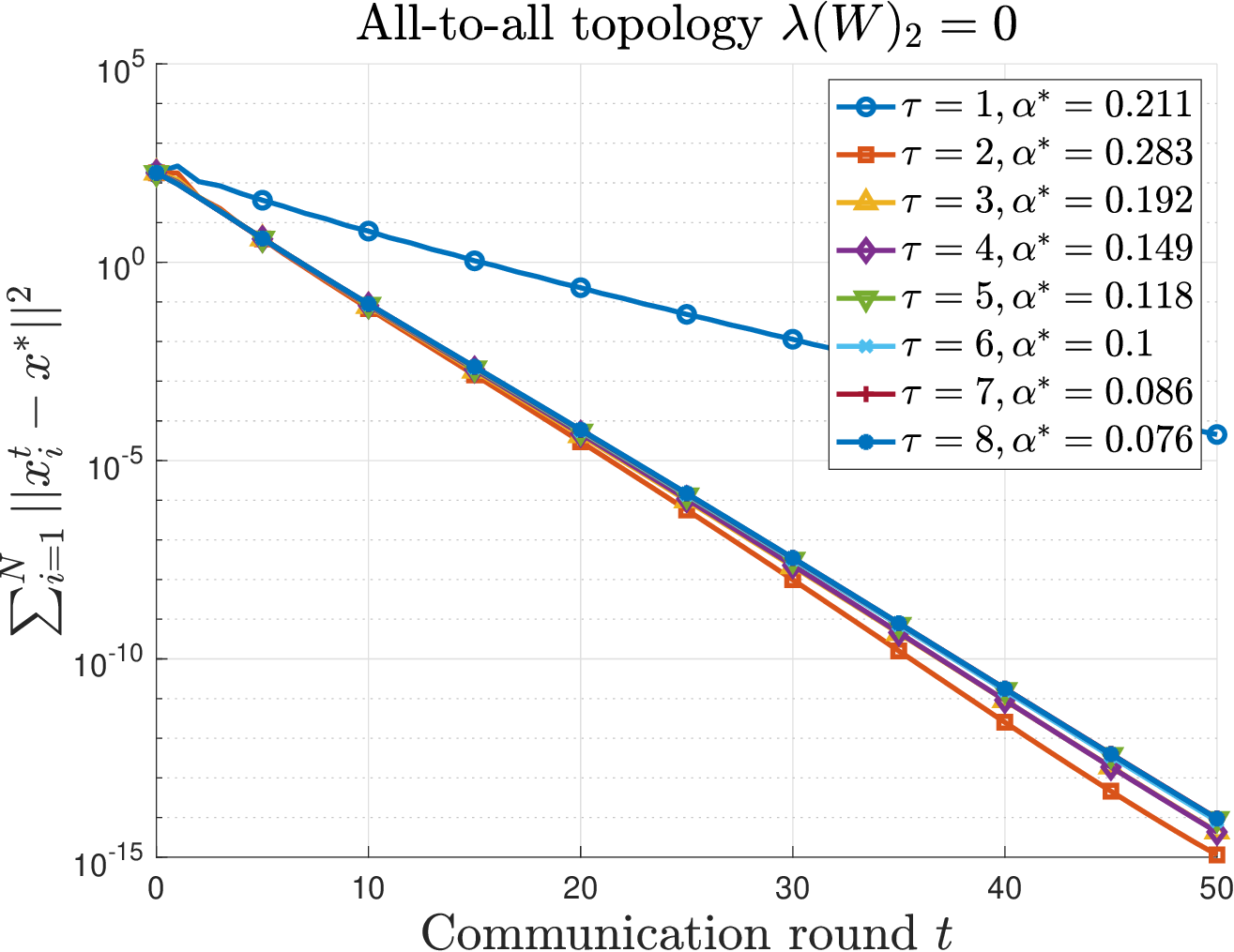}
    \caption{}
    \label{fig:alltoall_regression}
\end{subfigure}
\hfill
\begin{subfigure}{\columnwidth}
    \centering
    \captionsetup{font=small}
    \includegraphics[width=0.8\linewidth]{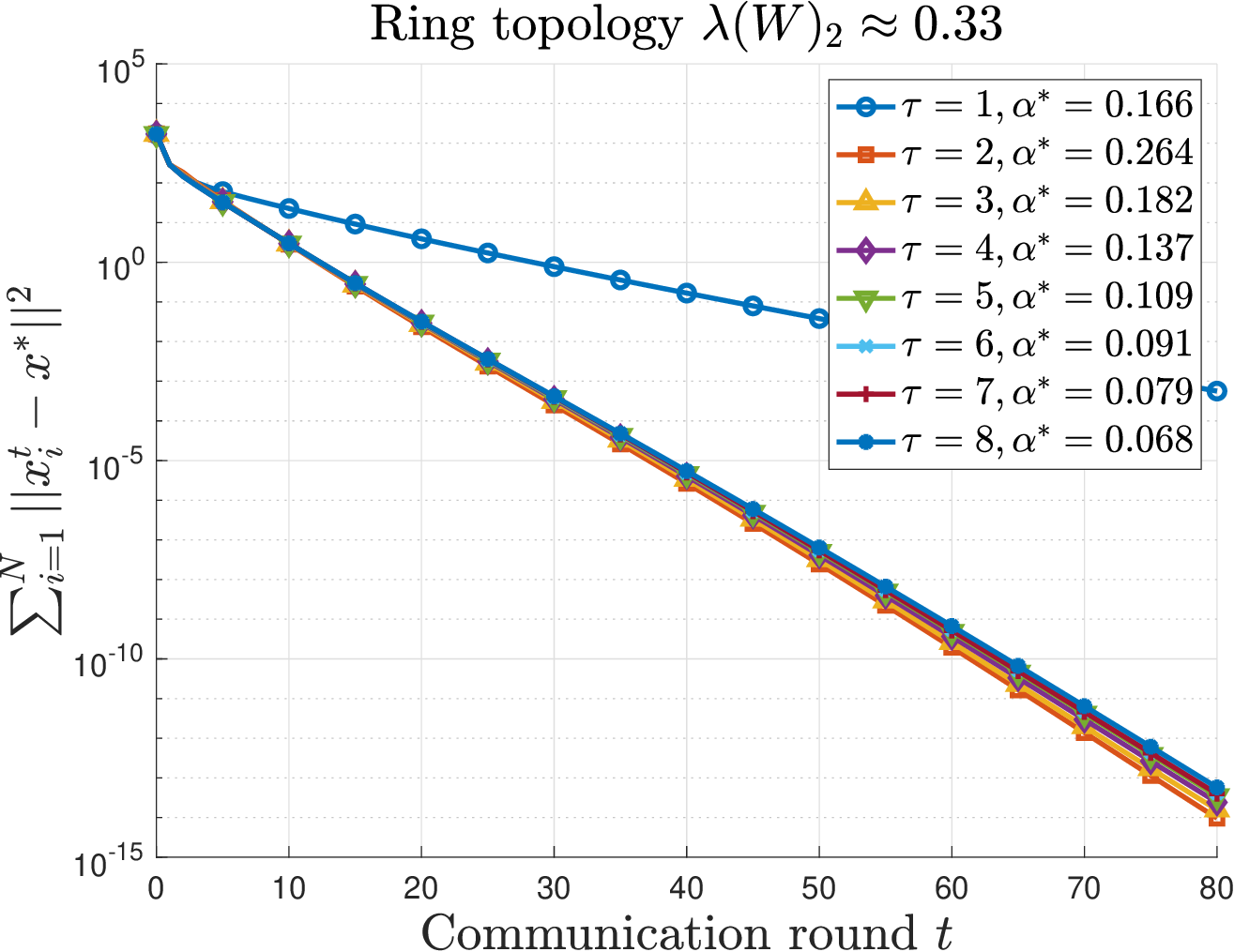}
    \caption{}
    \label{fig:ring_regression}
\end{subfigure}
\hfill
\begin{subfigure}{\columnwidth}
    \centering\
    \captionsetup{font=small}
    \includegraphics[width=0.8\linewidth]{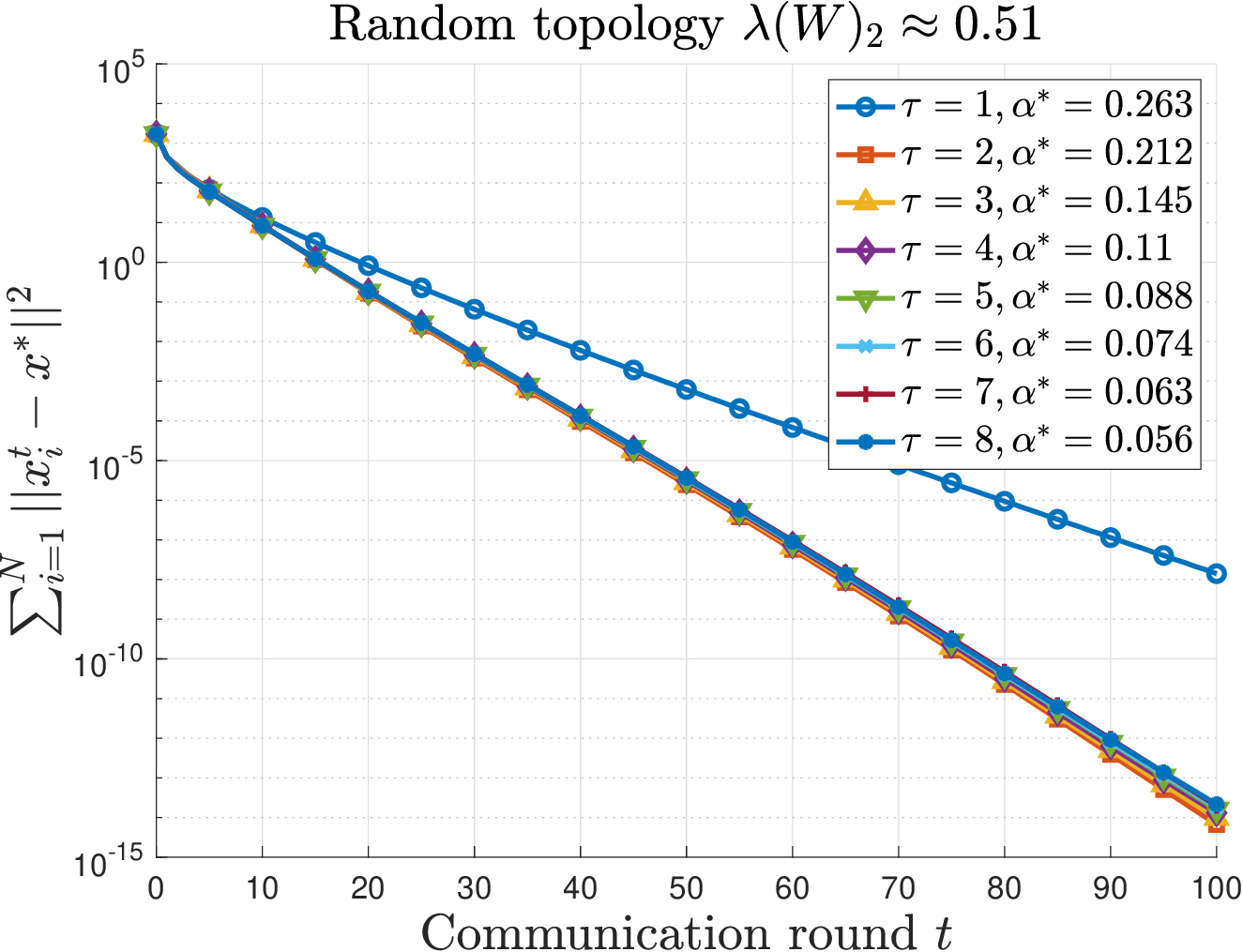}
    \caption{}
    \label{fig:random_regression}
\end{subfigure}
\begin{subfigure}{\columnwidth}
    \centering
    \captionsetup{font=small}
    \includegraphics[width=0.8\linewidth]{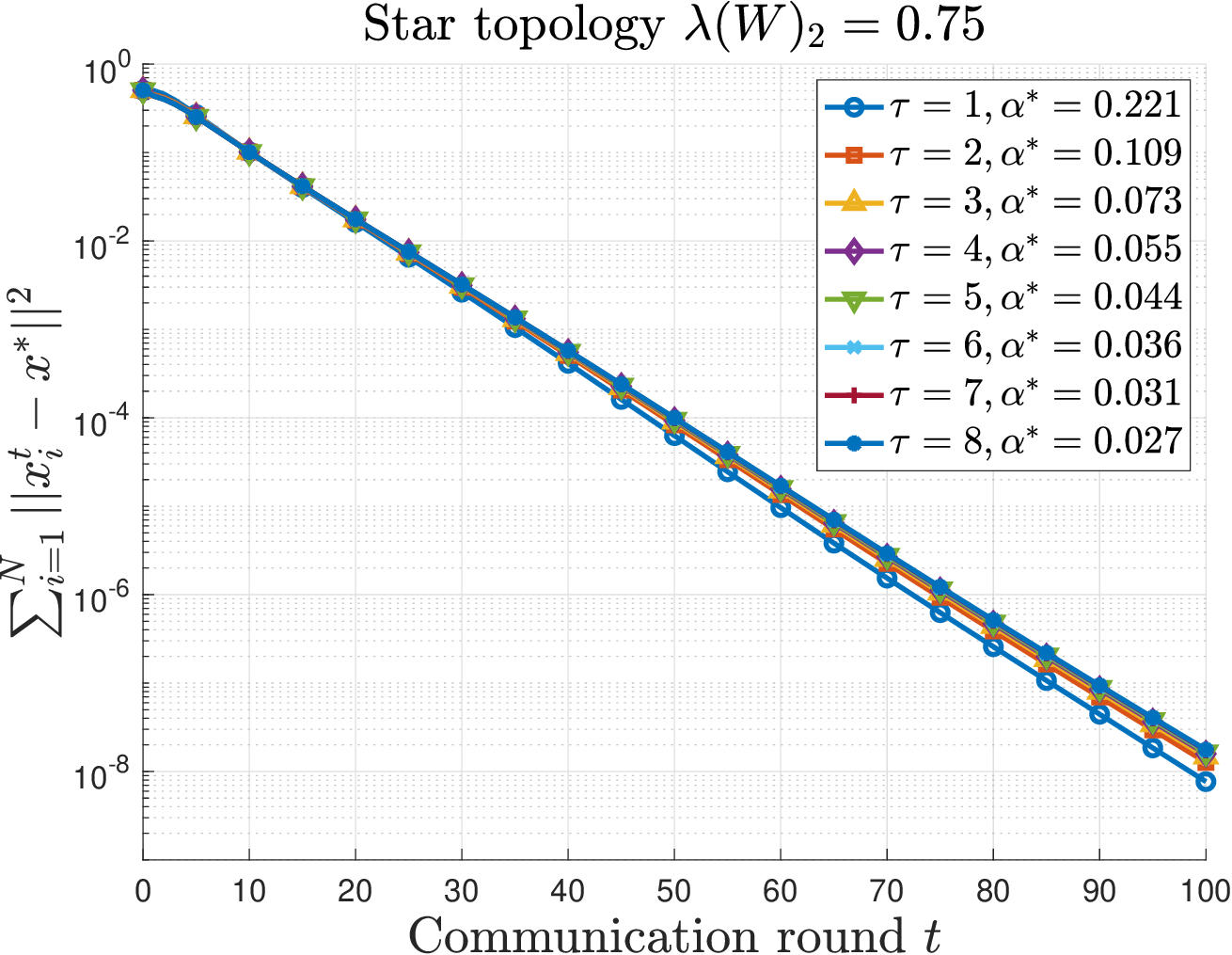}
    \caption{}
    \label{fig:star_regression}
\end{subfigure}
\caption{DIGing-based training of the linear regression model under different numbers of local updates. The network consists of 4 agents for each topology. The step size for each $\tau$ was determined via a grid search over $\alpha \in [0.01, 0.5]$ with 0.001 resolution. 
}
\label{fig:topology_regression}
\end{figure*}
\subsection{Regression}
\subsubsection{Problem Formulation} Linear regression is a supervised learning method. It models the relationship between a response variable and a set of input features, assuming a linear dependence. The model parameters are estimated by minimizing the mean squared error. Given local data \( A_i \in \mathbb{R}^{m \times d} \) and labels \( b_i \in \mathbb{R}^m \), we consider the distributed regression problem 
\[ \min_{w \in \mathbb{R}^d} \frac{1}{N}\sum_{i=1}^N f_i(w) = \frac{1}{N} \sum_{i=1}^N\|A_i w - b_i\|_2^2 .\] To ensure heterogeneity among randomly generated data samples, we leverage the Hessian matrix $ \nabla^2 f_i(\omega) = 2 A_i^T A_i $, which gives the Lipschitz constant as $ \lambda_{\max}(2 A_i^T A_i) $ and the strong convexity parameter as $ \lambda_{\min}(2 A_i^T A_i) $. Exploiting this property, we can generate random matrices $A_i$ and corresponding $b_i = A_i w_i^\star$ with desired Lipschitz smoothness parameters, strong convexity parameters, and distribution of local optima, thereby facilitating a controlled experimental setup.

\subsubsection{Parameter Settings} In our experiments, we consider \(m = 50 \) and \(d = 50\) with four different communication graphs. We initialize the iterates within the admissible set of global minimizers, following \cite{zamani2024set}, which can be achieved efficiently by having each agent minimize its local loss before participating in the distributed optimization process. 

\subsubsection{Results} 
\begin{figure}[ht]
    \centering
    \includegraphics[width=0.8\columnwidth]{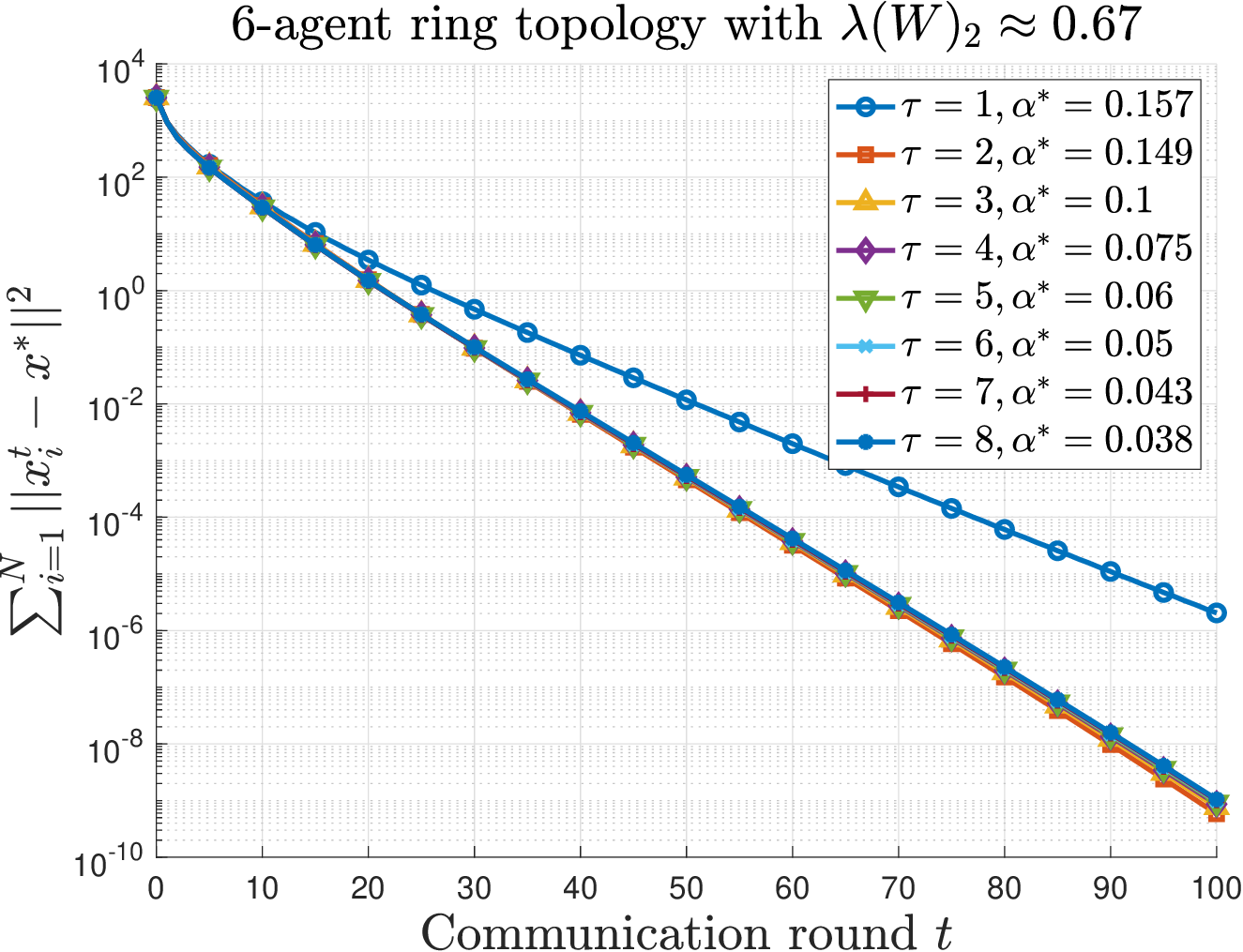}
    \caption{DIGing-based training of linear regression model under different numbers of local updates. The network consists of 6 agents arranged in a ring communication graph. The step size for each $\tau$ was determined via a grid search over $\alpha \in [0.01, 0.5]$ with 0.001 resolution.}
    \label{fig:6_ring}
\end{figure}

\begin{figure}[ht]
    \centering
    \includegraphics[width=0.8\columnwidth]{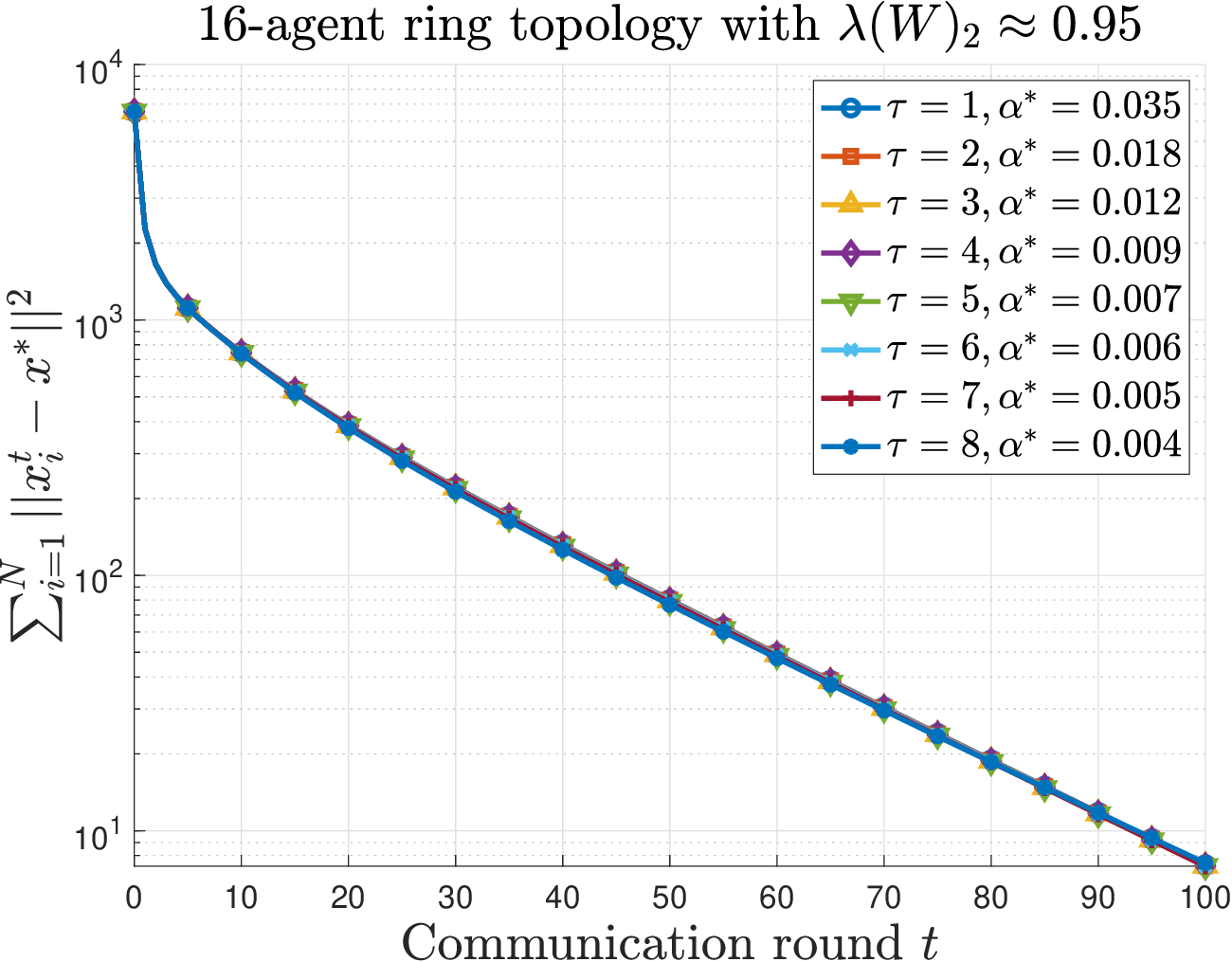}
    \caption{DIGing-based training of linear regression model under different numbers of local updates. The network consists of 16 agents arranged in a ring communication graph. The step size for each $\tau$ was determined via a grid search over $\alpha \in [0.01, 0.5]$ with 0.001 resolution.}
    \label{fig:16_ring}
\end{figure}

The results are presented in Figure~\ref{fig:topology_regression}, which shows that the maximal improvement is achieved when $\tau=2$, and the optimal step sizes for different values of $\tau$ follow the same pattern ($\alpha^{\ast} \propto 1/\tau$) as those confirmed by the PEP results. In Figure~\ref{fig:6_ring} and Figure~\ref{fig:16_ring}, we use the ring matrix with more than 4 agents. Note that as the number of agents increases, the ring graph exhibits progressively smaller spectral values. The results corroborate our theoretical predictions: the achievable speedup from multiple local updates decreases as the second largest eigenvalue of the mixing matrix becomes larger (i.e., the graph becomes sparser).
\subsection{Training of Convolutional Neural Networks (CNNs)}
\subsubsection{ CNN Setting}
We evaluate the impact of multiple local updates using a CNN on the MNIST dataset, considering both 10-agent all-to-all and ring graph topologies under highly heterogeneous data distributions, following the experimental setup in \cite{hsu2019measuring}. The architecture of the CNN consists of three convolutional layers with 32, 64, and 128 output channels, respectively, each utilizing a $5 \times 5$ kernel and a padding of 1. We employ the Softplus activation function throughout all hidden layers to ensure smoothness. Each convolutional stage is followed by an average pooling layer with a $2 \times 2$ window and a stride of 2. The resulting 512-dimensional latent feature vector is processed through two fully connected layers with 84 and 10 units, respectively. 
\subsubsection{Experiments Setting}
To isolate the effects of local updates from stochastic gradient noise, we use full-batch gradients throughout training. To determine the optimal learning rate, we perform an extensive grid search for the step size $\eta \in [0.01, 0.5]$ using an adaptive resolution: a fine-grained interval of 0.01 is utilized for smaller step sizes, with larger increments applied as the magnitude increases. The selection criterion is based on minimizing the cumulative loss over the final 50 communication rounds. For the initialization of local models, we adopt the approach proposed in \cite{zamani2024set}, in which agents perform preliminary local optimization before engaging in the global training process. 
\subsubsection{Results} 
\begin{figure}[ht]
    \centering
    \includegraphics[width=0.9\columnwidth]{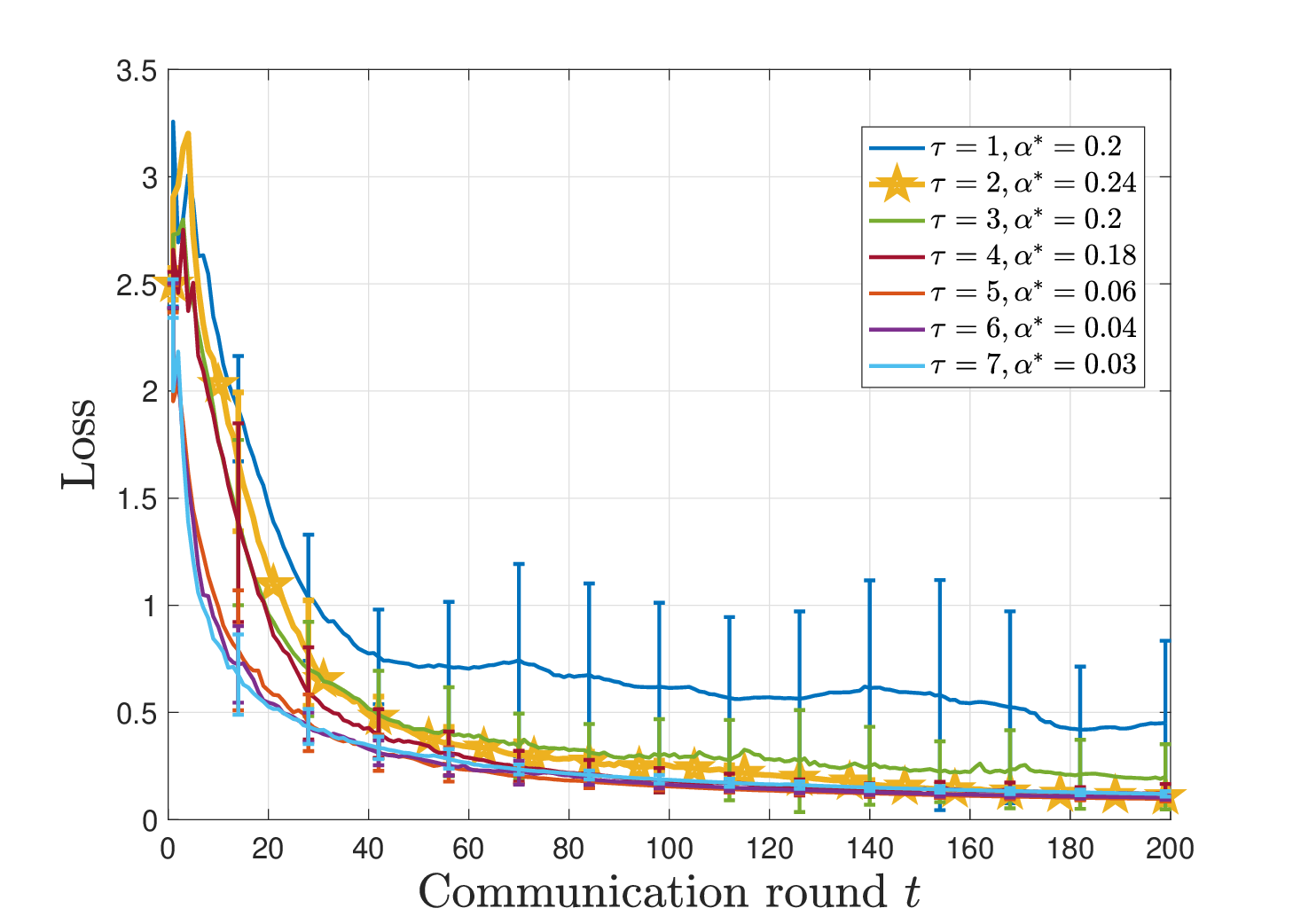}
    \caption{DIGing-based training of CNN on the MNIST dataset under different numbers of local updates. The network consists of 10 agents arranged in an all-to-all communication graph. The step size for each $\tau$ was determined via a grid search over $\alpha \in [0.01, 0.6]$. Each loss value is averaged over 5 runs, with error bars indicating the standard deviation.}
    \label{fig:mnist_alltoall}
\end{figure}

\begin{figure}[ht]
    \centering
    \includegraphics[width=0.9\columnwidth]{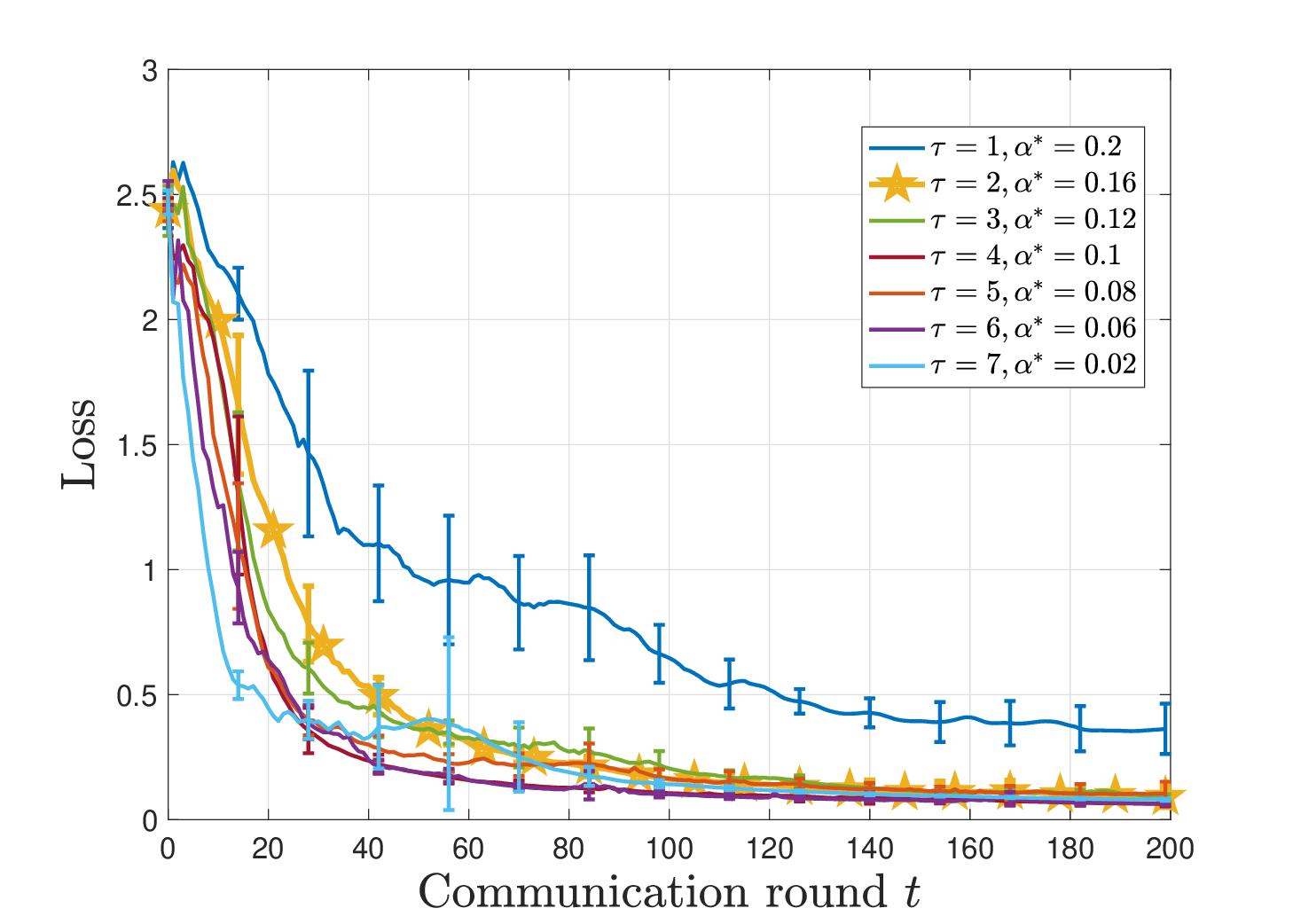}
    \caption{DIGing-based training of CNN on the MNIST dataset under different numbers of local updates. The network consists of 10 agents arranged in a random graph generated according to the Erdős--Rényi model with edge probability \(0.6\).  The step size for each $\tau$ was determined via a grid search over $\alpha \in [0.01, 0.5]$. Each loss value is averaged over 5 runs, with error bars indicating the standard deviation.}
    \label{fig:mnist_random}
\end{figure}

The results presented in Figures~\ref{fig:mnist_alltoall} and~\ref{fig:mnist_random} empirically validate our PEP-based theoretical findings. In particular, they show that the maximum improvement is achieved at $\tau = 2$, while additional local updates do not yield further gains, highlighting the robustness and applicability of our theoretical findings in practical learning scenarios.

\section{Conclusions}
Although local updates have been receiving increased attention in distributed optimization, existing results uniformly require the step size to decrease as the number of local updates increases. This makes it difficult to determine whether local updates can genuinely accelerate distributed optimization under exact gradients. In this paper, we leverage the exact performance bounds provided by the PEP framework to theoretically show that local updates can indeed accelerate convergence for a general class of functions. Interestingly, our PEP-based analysis indicates that performing just two local updates is sufficient to
achieve the maximal improvement when an appropriate step size is chosen. These findings are practically significant, as increasing the number of local updates also raises computational complexity. Additionally, we show that the convergence speed gain obtained from multiple local updates is affected by the network connectivity, with sparser
or less connected graphs—characterized by the spectral
properties of the mixing matrix—yielding smaller improvements.  Comprehensive experiments on real-world benchmark learning tasks confirm these theoretical results.

\appendices

\section{Proof of Theorem \ref{theorem_1}}\label{proof of theorem 1}
In this section, we prove Theorem \ref{theorem_1} by formulating the Performance Estimation Problem into an equivalent SDP. We use $\mathbb{S}_{+}$ to denote the set of all symmetric positive semidefinite matrices, and then we define
\begin{equation}
\begin{aligned}
&P = [{\bm x}^0,\; {\bm g}^0,\; {\bm g}^1,\; \dots,\; {\bm g}^K,\; {\bm g}^{\star},\; {\bm g}^{\ast},\; {\bm x}^{\star},\; {\bm x}^{\ast}]
   \in \mathbb{R}^{d\times [(K+6)N]}, \\
&G = P^{\top}P \in \mathbb{S}^{((K+6)N)\times((K+6)N)}_{+},\notag\\
&F = [{\bm f}^0,\; {\bm f}^1,\; \dots,\; {\bm f}^K,\; {\bm f}^{\star},\; {\bm f}^{\ast}] \in \mathbb{R}^{1\times(K+3)N},
\end{aligned}
\end{equation}
$G$ and $F$ are the variables of the resulting SDP from the PEP. We use the standard notation \( e_i \in \mathbb{R}^d \) to denote the unit \(d\)-dimensional vector with the only nonzero entry ``1''  in the \( i \)-th component. The following vectors are employed to select specific columns in \(P\) and \(F\):
\begin{equation*}
\begin{aligned}
&\textbf{f}_i^{k} = e_{kN+i}\in \mathbb{R}^{(K+3)N},\textbf{f}_i^{\star} =  e_{(K+1)N+i} \in \mathbb{R}^{(K+3)N},\\
&\textbf{f}^{\ast}_{i} = e_{(K+2)N+i}\in\mathbb{R}^{(K+3)N},\textbf{g}_i^{k} = e_{(k+1)N+i}\in \mathbb{R}^{(K+6)N},\\
&\textbf{g}_i^{\star} =e_{(K+2)N+i}\in \mathbb{R}^{(K+6)N},\textbf{g}_{i}^{\ast} =e_{(K+3)N+i}\in \mathbb{R}^{(K+6)N},\\
&\textbf{x}_i^{0} = e_{i}\in \mathbb{R}^{(K+6)N},
\textbf{x}^{\star}_i = e_{(K+4)N+i}\in \mathbb{R}^{(K+6)N},\\
&\textbf{x}_i^{\ast} = e_{(K+5)N+i}\in \mathbb{R}^{(K+6)N},\forall i\in \mathcal{S}, k \in I_K.\\
\end{aligned}
\end{equation*}
The stacked version of these vectors is analogous to the formulation in Section \ref{notations} and is omitted here for the sake of brevity. Now we are in a position to give our formulation of the PEP for a decentralized algorithm $\mathcal{M}$ as follows
\begin{align}
\max_{F,G} \quad & \quad  \langle G, (\sum_{i=1}^N (\textbf{x}^K_i - \textbf{x}_i^*))(\sum_{i=1}^N (\textbf{x}^K_i - \textbf{x}_i^*)^T)\rangle\label{sdp_performance_measure}\\
\text{subject to}
\quad & \quad \langle G, A(\{\textbf{x}_{i}^k, \textbf{g}_{i}^k, \textbf{f}_i^k\}_{k\in \{p,q\}}, \mu, L)\rangle \notag\\
\quad &\quad \le F*(\textbf{f}_i^{p} - \textbf{f}_i^{q}), \;  \forall i \in \mathcal{S},\; p, q \in I_K^{\star,\ast},\label{sdp_interpolation_constraints}\\
\quad & \quad \{\textbf{x}_{i}^k\}_{i\in \mathcal{S}, k\in I_K} \;\text{are generated recursively}\notag\\
\quad & \quad\text{by algorithm}\;\mathcal{M},\label{sdp_algorithm_constraints}\\
\quad & \quad \langle G\;,(\textbf{x}^{*}_1 - \textbf{x}^*_i)(\textbf{x}_1^* - \textbf{x}^{\ast}_i)^T\rangle = 0,\;\forall i\in\mathcal{S},\label{identical_all_gloal_optimum}\\
\quad & \quad \langle G\;,(\sum_{i=1}^N\textbf{g}^{*}_i)(\sum_{i=1}^N\textbf{g}_i^{*})^T\rangle = 0,\;\forall i \in \mathcal{S},\notag\\
\quad & \quad\langle G, \textbf{g}_i^{\star}\textbf{g}_i^{\star T}\rangle=0 ,\;\forall i \in \mathcal{S},\label{sdp_optimal_values}\\
\quad & \quad \langle G, (\textbf{x}_i^0-\textbf{x}^*_i)(\textbf{x}_i^0-\textbf{x}^*_i)^T\rangle \le R_0^2,\;\forall i\in \mathcal{S},\label{sdp_initial_constraint}\\
\quad & \quad \langle G, (\textbf{x}_i^{\star}-\textbf{x}^*_i)(\textbf{x}_i^{\star}-\textbf{x}^*_i)^T\rangle \le R_*^2,\;\forall i \in \mathcal{S},\label{sdp_optimum_constraint}\\
\quad & \quad G \succeq 0,\label{sdp_semidefinite_constraint}\\
\quad & \quad \mathrm{rank}(G) \le d,\label{sdp_rank_G_constraint}
\end{align}
where
\begin{align*}
    &A(\{\textbf{x}_{i}^k, \textbf{g}_{i}^k, \textbf{f}_i^k\}_{k\in \{p,q\}}, \mu, L)\\
    = &\frac{1}{2}\left[(\textbf{x}^q_i - \textbf{x}^p_i){\textbf{g}^{q}_i}^T + \textbf{g}^{q}_i(\textbf{x}^q_i - \textbf{x}^p_i)^T\right]\\
    &- \frac{\mu(L-\mu)}{2L^2} \left[(\textbf{x}^q_i - \textbf{x}^p_i)(\textbf{g}^q_i - \textbf{g}^p_i)^T\right]\\
    &-\frac{\mu(L-\mu)}{2L^2}\left[(\textbf{g}^q_i - \textbf{g}^p_i)(\textbf{x}^q_i - \textbf{x}^p_i)^T\right]\\
    &+\frac{1}{2} \left( 1 - \frac{\mu}{L} \right)\left[\frac{1}{L}  (\textbf{g}^p_i - \textbf{g}^q_i)(\textbf{g}^p_i - \textbf{g}^q_i)^T\right]\\
    &+\frac{1}{2} \left( 1 - \frac{\mu}{L} \right)\left[\mu (\textbf{x}_i^p - \textbf{x}_i^q )(\textbf{x}_i^p - \textbf{x}_i^q)^T \right],
\end{align*}
and \(\langle\cdot, \cdot\rangle\) denotes the standard matrix inner product defined as \(\langle A, B\rangle\) = \(\mathrm{trace}(AB^T)\).
(\ref{sdp_performance_measure}) represents the measure $\sum_{i=1}^N \|x_i^K - x^*\|^2$. Constraints (\ref{sdp_interpolation_constraints}) are the necessary and sufficient conditions for smooth, strongly convex interpolation; that is, the points indexed by 
 \(I\) can be interpolated by a function satisfying all constraints in (\ref{sdp_interpolation_constraints}) for all \(i, j\in I\). So the infinite dimensional function constraint \(f_i \in \mathcal{F}_{\mu, L}\) can now be expressed in terms of all the iterate and optimal points \(\{(x_{i}^k, g_i^k, f_i^k)\}_{i\in \mathcal{S},k\in I_K^{\star, \ast}}\)only, using the interpolation conditions given above.
Constraints (\ref{sdp_interpolation_constraints}) corresponds to the interpolation constraints of Lemma (\ref{lemma}), where we use the simple identity
    \(
        \langle g_j, x_i - x_j \rangle = \frac{1}{2}\langle g_j, x_i - x_j \rangle + \frac{1}{2}\langle x_i - x_j, g_j \rangle.
    \)
Consequently, the resulting Gram matrix expression
\begin{equation}
\begin{aligned}
    &\frac{1}{2}\textbf{g}_j^\top P^\top P (\textbf{x}_i - \textbf{x}_j) + \frac{1}{2}(\textbf{x}_i - \textbf{x}_j)^\top P^\top P \textbf{g}_j\\
    = &\left\langle G, \frac{1}{2}\left((\textbf{x}_i - \textbf{x}_j)\textbf{g}_j^\top + \textbf{g}_j(\textbf{x}_i - \textbf{x}_j)^\top\right)\right\rangle\notag
\end{aligned}
\end{equation}
holds, where the matrix \(\frac{1}{2}\left((\textbf{x}_i - \textbf{x}_j)\textbf{g}_j^\top + \textbf{g}_j(\textbf{x}_i - \textbf{x}_j)^\top\right)\) is symmetric. This technique enables us to reformulate Lemma (\ref{lemma}) into the form of (\ref{sdp_interpolation_constraints}) for all points generated during the iteration process. Constraint (\ref{sdp_algorithm_constraints}) arises from the fact that, for each agent \(i\), we have the following relation:
    \(
        \textbf{x}_i^K \in \text{Span}\left\{\{\textbf{x}_{i}^k, \textbf{g}_{i}^k, \textbf{f}_i^k\}_{k \in I_K,\, i \in \mathcal{S}}\right\},
    \)
meaning that \(\textbf{x}_i^K\) is a linear combination of the elements \(\{\textbf{x}_{i}^k, \textbf{g}_{i}^k, \textbf{f}_i^k\}_{k \in I_K,\, i \in \mathcal{S}}\) generated by the algorithm \(\mathcal{M}\). Therefore, once the initial point is fixed, we can recursively determine \(\textbf{x}_i^K\) for all \(i \in \mathcal{S}\). Constraints (\ref{sdp_optimal_values}) ensure that \(x^*\) is one of the global optimum and \(x^{\star}_i\) is the local optimum of each \(f_i\). For constraints (\ref{sdp_initial_constraint}) and (\ref{sdp_optimum_constraint}), \( R_* \) bounds the distance between local and global optima, i.e., \( \|x_i^{\star} - x^*\| \leq R_* \), \( R_0 \) bounds the distance between the initial point and the global optimum, i.e., \( \|x^0_i - x^*\| \leq R_0 \).

According to \cite{pepsdp}, the optimization problem characterized by \eqref{performance_measure}-\eqref{last_two}, along with its
solution, is independent of the dimension $d$. Therefore, the rank constraint \eqref{sdp_rank_G_constraint} is redundant and can
be omitted. After removing \eqref{sdp_rank_G_constraint}, the optimization problem characterized by \eqref{performance_measure}-\eqref{last_two} becomes an
SDP.


\bibliographystyle{IEEEtran}
\bibliography{reference.bib}

\vspace{-1.2cm}


	

\end{document}